\documentclass[10pt]{article}
\usepackage{authblk}
\usepackage{amsmath}
\usepackage{array}
\usepackage{url}
\usepackage{xcolor}
\usepackage{graphicx}
\usepackage[a4paper, total={6in, 9in}]{geometry}

\begin{document}

\title{A Biomathematical Model of Tumor Response to Radioimmunotherapy with $\alpha$PDL1 and $\alpha$CTLA4}

\author[1, 2]{Isabel~Gonz\'{a}lez-Crespo}
\author[3]{Antonio~G\'{o}mez-Caama\~{n}o}
\author[1, 2]{\'{O}scar~L\'{o}pez~Pouso}
\author[4]{John~D.~Fenwick}
\author[1, 5, \thanks{juan.pardo.montero@sergas.es}]{Juan~Pardo-Montero}

\affil[1]{Group of Medical Physics and Biomathematics, Instituto de Investigaci\'{o}n Sanitaria de Santiago}
\affil[2]{Department of Applied Mathematics, Universidade de Santiago de Compostela, Spain} 
\affil[3]{Department of Radiation Oncology, Clinical University Hospital of Santiago, Spain}
\affil[4]{Department of Molecular and Clinical Cancer Medicine, Institute of Translational Medicine, University of Liverpool, Liverpool, United Kingdom}
\affil[5]{Department of Medical Physics, Clinical University Hospital of Santiago, Spain}

\date{}

\maketitle

\begin{abstract}
	There is evidence of synergy between radiotherapy and immunotherapy. Radiotherapy can increase liberation of tumor antigens, causing activation of antitumor T-cells. This effect can be boosted with immunotherapy. Radioimmunotherapy has potential to increase tumor control rates. Biomathematical models of response to radioimmunotherapy may help on understanding of the mechanisms affecting response, and assist clinicians on the design of optimal treatment strategies. In this work we present a biomathematical model of tumor response to radioimmunotherapy. The model uses the linear-quadratic response of tumor cells to radiation (or variation of it), and builds on previous developments to include the radiation-induced immune effect. We have focused this study on the combined effect of radiotherapy and $\alpha$PDL1/$\alpha$CTLA4 therapies. The model can fit preclinical data of volume dynamics and control obtained with different dose fractionations and $\alpha$PDL1/$\alpha$CTLA4. A biomathematical study of optimal combination strategies suggests that a good understanding of the involved biological delays, the biokinetics of the immunotherapy drug, and the interplay between them, may be of paramount importance to design optimal radioimmunotherapy schedules. Biomathematical models like the one we present can help to interpret experimental data on the synergy between radiotherapy and immunotherapy, and to assist in the design of more effective treatments.
\end{abstract}

\section{Introduction}

Cancer immunotherapy (IT) is a therapeutic strategy against cancer that aims at boosting and exploiting the natural immune response to control and cure tumors \cite{whiteside2016, waldmann2003}.
Checkpoint inhibitors are a type of IT which is used for treatment of several cancers, including melanoma, prostate, NSCLC and leukemia \cite{kwon2014, golden2013}. These inhibitors block different checkpoint proteins, like CTLA-4 and PD-1/PD-L1, which are well known suppressors of the immune response against tumors. Inhibitors of these proteins have shown promising results in preclinical experiments, and there are several monoclonal antibodies against PD-1/PD-L1 approved to treat different types of cancer \cite{gong2018}. However, efficacy of IT as cancer treatment is still limited, but for particular cases. For example, ipilimumab has shown an improvement on the survival of melanoma patients, but response rates are low, in the 10-15\% range \cite{hodi2010}.

Many preclinical studies have shown that the combination of radiotherapy (RT) and IT, in particular inhibitors of CTLA-4, and PD-1/PD-L1, is significantly more effective than RT and IT alone \cite{herrera2016, dewan2009, deng2014, sperduto2015, crittenden2015}. The dominant biological cell killing mechanism behind the effect of RT is the generation of double strand breaks in the DNA by ionizing particles \cite{mladenov2013}. However, there is evidence that radiation can trigger other cell killing mechanisms, particularly when delivered at high-doses per fraction, which may be important for the synergy with immunotherapy. High-doses of radiation can damage the tumor vascular system, eventually triggering cell death \cite{song2015}, and also lead to an immune response against surviving tumor cells \cite{sperduto2015, song2012, cruzMerino2014}, increasing the likelihood of tumor control. The mechanisms behind these induced immune effects seem to be related to the increased liberation of tumor antigens, which cause the activation of antitumor T-cells, and the modification of tumor microenvironment, killing immune down-regulators like Treg and MDSC cells, and facilitating T-cell infiltration in the tumor \cite{herrera2016, chen2013, walle2018}.

Despite the potential of IT and RT, how the best combination of both therapies can be achieved is still a matter of study. In this regard, validated biomathematical models of RT+IT (\textit{in silico} tumor models) would be very useful. Biomathematical models, based on solid experimental and clinical data are of high importance in order to interpret results, and may also assist on the design of optimal therapeutic strategies, potentially guiding clinicians in the selection of optimal treatments. Modeling the response of tumors to IT has been addressed in the biomathematical literature, following both phenomenological and systems biology approaches \cite{depillis2005, depillis2006, chareyron2009, lai2017, radunskaya2018, benchaib2019, eftimie2011}. On the other hand, modeling the synergistic combination of IT and RT has been less studied due to the novelty of this treatment strategy, but it is becoming an active field of research \cite{serre2016, serre2018, kosinsky2018, poleszczuk2018, sung2020, butner2020}. We would like to highlight the recent works of Serre et al. and Kosinsky et al. \cite{serre2016, serre2018, kosinsky2018}. In the former paper, a simple model of response to radioimmunotherapy with inhibitors of CTLA-4 and PD-L1 is presented and used to fit experimental data of tumor response and rejection probability of implanted tumors. The latter study presents a more complicated model, based on ordinary differential equations, which is used to fit experimental data and to formulate hypothesis of optimal treatment strategies with RT+inhibitors of PD-L1.

In this article we build upon these works to develop a dynamical model of tumor response to radioimmunotherapy with inhibitors of PD-L1 ($\alpha$PDL1) and CTLA-4 ($\alpha$CTLA4). The model presented consists in a system of coupled impulsive delay differential equations (DDE) that describe the time evolution of the most relevant cell populations. DDE are used to explicitly include biological response delays, which may have importance when modeling biological systems. The model is presented both as a continuous deterministic model and as a discrete stochastic \textit{Markov-birth-death} model (the latter formulation is especially important when modelling tumor control achieved with a given treatment). The model has been tested by fitting experimental preclinical data of tumor response to combined therapies of radiation with different fractionations and  $\alpha$PDL1/$\alpha$CTLA4 \cite{dewan2009, deng2014}, including volume dynamics and tumor control.

\section{Methods and Materials}

\subsection{Overview of the Model}

\begin{figure}[!t]
	\centering
	\includegraphics[width=4.2in]{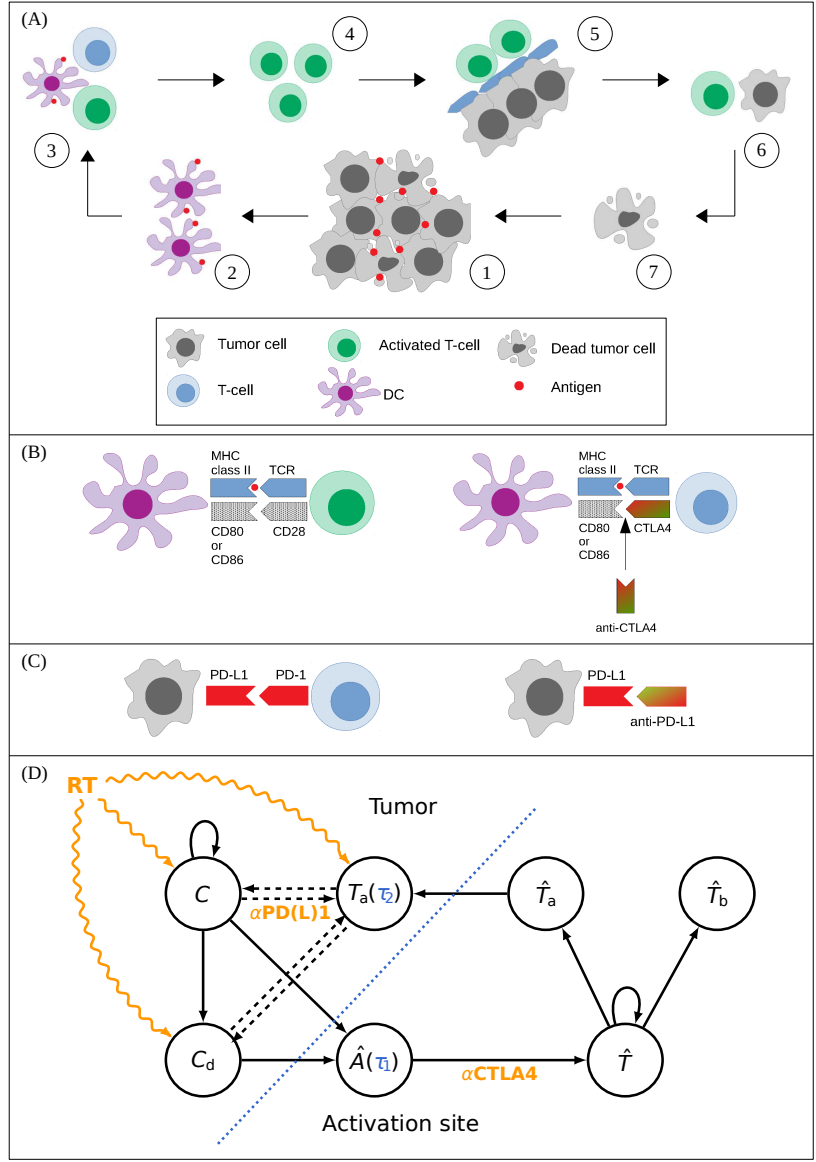}
	\caption{Sketch of the tumor immunity cycle and the role of $\alpha$PDL1/$\alpha$CTLA4, that serves as the basis for our model, and sketch of the flowchart of the biomathematical model. (A) Sketch of the tumor immunity cycle: 1) Tumor cells release antigens, either naturally or responding to radiotherapy damage; 2) dendritic cells (DC) take antigens to activation sites (lymph nodes); 3) T-cells against tumor cells are activated; 4) T-cells migrate to the tumor and 5) infiltrate it; 6) T-cells attack tumor cells and 7) kill them. (B) Sketch of T-cell activation blocking by the CTLA-4 receptor, and the potential role of $\alpha$CTLA4. (C) Sketch of the inactivation of T-cells through the PD-L1 receptor, and role of $\alpha$PDL1 in avoiding such inactivation. (D) Sketch of the flowchart of the model showing interconnections between different compartments: viable tumor cells (${{C}}$), doomed tumor cells (${{C}}_\mathrm{d}$), active T-cells in the tumor (${{T_\mathrm{a}}}$), antigens/APCs (${{\hat{A}}}$), as well as pool of T-cells (${{\hat{T}}}$), activated T-cells (${{\hat{T}_\mathrm{a}}}$) and blocked T-cells (${{\hat{T}_\mathrm{b}}}$).  The dashed line shows the separation between compartments physically located in the tumor (left) or in the activation sites (right). The notation also distinguishes between the two locations, identifying the compartments within the activation zone with a hat `` $\hat{}$ ''. $\tau_1$ and $\tau_2$ are biological delays for migration between those two physical locations. The effect of radiotherapy (RT), and $\alpha$PDL1/$\alpha$CTLA4 is also indicated.}
	\label{fig_cycle}
\end{figure}

Our biomathematical model follows the cancer-immunity cycle \cite{chen2013}, presented in Fig.~\ref{fig_cycle}(A) together with sketches of the role of $\alpha$CTLA4, Fig.~\ref{fig_cycle}(B), and  $\alpha$PDL1, Fig.~\ref{fig_cycle}(C). A sketch of the biomathematical model is shown in Fig.~\ref{fig_cycle}(D). Tumor cells release antigens, naturally and due to cell death; antigens are taken by antigen-presenting-cells (APCs) to activation sites (lymph nodes) where they trigger T-cell activation; active T-cells migrate to the tumor, infiltrate it, and kill tumor cells. Each radiation dose fraction kills tumor cells (contributing to the release of antigens), as well as T-cells present in the tumor. Immunotherapy drugs affect either the de-inhibition of T-cells ($\alpha$CTLA4) or the tumor cell-killing by activated T-cells ($\alpha$PDL1). The model includes three main species: tumor cells, T-cells, and antigens/APCs. These species are split into six different compartments, which exist in two different spatial locations: in the tumor (viable {cancer} cells, {$C$};  tumor cells that are doomed by radiation and will eventually die, {$C_\mathrm{d}$}; active T-cells against tumor cells, {$T_\mathrm{a}$}), and in the activation sites (a phenomenological compartment accounting for the effects of antigens and APCs, {$\hat{A}$};  the T-cell pool, {$\hat{T}$}; activated T-cells, {$\hat{T}_\mathrm{a}$}; blocked T-cells, {$\hat{T}_\mathrm{b}$}). The compartments at the activation sites are identified with a hat ( $\hat{}$ ) to facilitate the understanding of the model. Migration between different spatial locations (antigens/APCs migrating from the tumor to activation sites and activated T-cells migrating from the activation sites and infiltrating the tumor) result in biological delays which are explicitly included in our model by using delay differential equations. The effect of radiation dose delivery is treated as an impulse (thus, formally, the differential equations are impulsive). The main terms involved in the dynamics of our model can be shown in a simple way as:
\begin{align*}
	\text{dynamics }&C\text{ = proliferation - radiation death - immune death}\\
	\text{dynamics }&\hat{A}\text{ = natural release + {RT}-mediated release - natural elimination - T-cell activation}\\
	\text{dynamics }&T_\mathrm{a}\text{ = activation/infiltration - radiation death - immune death - natural elimination}
\end{align*}

The model is built following a mechanistic approach rather than a purely phenomenological approach. It includes many simplifications, though, in order to present a tractable problem. Among the most relevant simplifications, we should cite:  i) We do not include spatial coordinates in our multi-compartmental model; ii) The process of immune death is overly simplified. In particular, we do not include other cell types that participate in this process, either favoring immunity or acting as suppressors, like natural killers or T-regs; iii) We mostly rely on the linear-quadratic model (LQ) to account for radiation cell death, but departures from the LQ model at high-doses are studied and discussed; iv) The process of T-cell infiltration in the tumor is not included in the model.

\subsection{Direct Cell Death and Kinetics}
\label{section:lq}

Dose delivery is modeled as instantaneous (it seems like a good approximation, as in typical fractionated treatments dose delivery takes minutes, while the typical times of our model are days). Radiation tumor cell death is typically modeled with the linear-quadratic (LQ) model:
\begin{equation}
	\label{eq_LQ}
	\log{\mathit{SF}} = -\alpha d - \beta d^2
\end{equation}
where $\mathit{SF}$ is the surviving fraction of a population of cells after being irradiated to a radiation dose, $d$, and $\alpha$ and $\beta$ are the LQ linear and quadratic parameters.

We will fit data with different fractionation schedules. It is well known that cell death can depart from the standard LQ-model, especially at high doses per fraction. Several effects contribute to this, including re-oxygenation~\cite{wouters1997}, saturation \cite{kirkpatrick2008}, or vascular damage \cite{song2015}. In order to investigate possible departures from the LQ model, we will check whether other models provide better fits. In particular, we will investigate a simple ad hoc modification of the quadratic term of the LQ-model~\cite{gago2021}:
\begin{equation}
	\label{eq_LQBetaMod}
	\beta \rightarrow \beta_0 \left(1+c\sqrt{d}\right)
\end{equation}
where $c$ is a free parameter and $d$ is the dose per fraction.

We will also use the Linear-Quadratic-Linear (LQL) model \cite{guerrero2004}, which obtains the surviving fraction as:
\begin{equation}
	\label{eq_LQL}
	\log{\mathit{SF}} = -\alpha d - 2 \beta \frac{ xd-1+e^{-´xd}}{x^2}
\end{equation}
where $x$ is an extra parameter that modulates the slope of the curve.

We will use (\ref{eq_LQ}), (\ref{eq_LQBetaMod}) and (\ref{eq_LQL}) to model radiation-induced tumor cell death. T-cell death will be described using the LQ-model.

Cells fatally damaged by radiation (doomed) do not die instantly, but follow a given kinetics, generally a pause (mitotic delay) followed by a progressive death as lethally damaged cells enter mitosis and suffer mitotic catastrophe. We will model this by considering a mitotic delay followed by an exponential death \cite{gago2016}. 

Viable tumor cells can proliferate, and we describe this by using the logistic formalism \cite{fedotov2008}. On the other hand, while doomed cells may carry some proliferative capacity (abortive divisions \cite{dorr1997}), it should be limited and does not contribute to the long-term cell population. Therefore, we will ignore it. 

From the above considerations we can write the following equation for viable tumor cells:
\begin{equation}
	\label{eq_tumorCells}
	\frac{d{C}}{dt}(t)=\lambda_1 {C}(t) [1-\lambda_2C_\mathrm{tot}(t)] - K_{{\mathrm{C}}}(t)
\end{equation}
where $\lambda_1$ and $\lambda_2$ are constants and $C_\mathrm{tot}(t) = {{C}}(t)+{{C}}_\mathrm{d}(t)$ is the total number of tumor cells at time~$t$. $K_{{\mathrm{C}}}(t)$ is an impulse term accounting for the effect of the radiation dose:

\begin{equation}
	\label{eq_impulse}
	K_{{\mathrm{C}}}(t) = (1-\mathit{SF}_{{\mathrm{C}}}(d(t))) {C}(t) \sum_{i} \delta(t-t_i)
\end{equation}
where $\{t_i\}$ is the vector of radiation delivery times, $\{d_i\}$ are the doses delivered at the times $\{t_i\}$, and $\mathit{SF}_{\mathrm{C}}$ is the surviving fraction of tumor cells given by (\ref{eq_LQ}), (\ref{eq_LQBetaMod}) or (\ref{eq_LQL}). $\delta(x)$~is the Dirac delta function.

We consider that each radiation fraction creates new doomed cells, but does not interfere with the radiation kinetics of existing ones. Therefore, we can split the compartment of doomed cells into $n$ compartments created by $n$ dose fractions:
\begin{align}
	\label{eq_doomedCells1}
	\frac{d{{C}}_{\mathrm{d},i}}{dt}(t)&= K_{{{\mathrm{C}}}}(t) - \phi \omega(\bar{t}_i) {{C}}_{\mathrm{d},i}(t)\\	
	\label{eq_doomedCells2}
	{{C}}_\mathrm{d}(t)&= \sum_{i} {{C}}_{\mathrm{d},i}
\end{align}
Here, ${{C}}_{\mathrm{d},i}(t)$ denotes doomed cells created by the radiation dose fraction $d_i$, $t_i$ is the delivery time of that fraction, and $\bar{t}_i = t-t_i$. Notice that ${{C}}_{\mathrm{d}´,i}(t)$ is defined as zero for $t<t_i$. The parameter $\phi$ is the death rate, and $\omega$ models the mitotic delay and progressive incorporation of damaged cells to cell death kinetics after a radiation fraction:
\begin{equation}
	\label{eq_doomedKinetic}
	\setlength{\nulldelimiterspace}{0pt}
	\omega(\bar{t})=\begin{cases}
	0,&\ \text{for }\bar{t} \leq {\tau_\mathrm{d}}_1\\
	\dfrac{\bar{t}-{\tau_\mathrm{d}}_1}{{\tau_\mathrm{d}}_2-\tau_{\mathrm{d}1}},&\ \text{for }{\tau_\mathrm{d}}_1 < \bar{t} \leq {\tau_\mathrm{d}}_1\\
	1,&\ \text{for }\bar{t} > {\tau_\mathrm{d}}_2
	\end{cases}
\end{equation}

Radiation also kills T-cells present in the tumor as:
\begin{equation}
	\label{eq_lymphocytes}
	\frac{d{{T_\mathrm{a}}}}{dt}(t)=-K_{{\mathrm{T}}}(t)
\end{equation}
where $K_{{\mathrm{T}}}(t)$ is the impulse term for T-cells, which has the same form of (\ref{eq_impulse}), but with $\mathit{SF}_{{\mathrm{T}}}$ as the surviving fraction of T-cells. It is assumed that radiation-damaged T-cells die instantly, and so there are no kinetic terms associated to such process.

\subsection{Antigen Release and T-cell Activation}
\label{section:antigen_activation}

Antigens are considered to be released both naturally (rate proportional to the number of tumor cells, both viable and doomed), and during radiation-induced cell death (proportional to the rate of cell death). We also include a term describing natural elimination. A biological delay, $\tau_1$, between antigen release and T-cell activation is included (which may be interpreted as the time that APCs take to collect antigens and carry them to the activation sites, Fig. \ref{fig_cycle}(A)):
\begin{equation}
	\label{eq_antigens}
	\frac{d{{\hat{A}}}}{dt}(t)=\rho {{C_\mathrm{tot}}}(t-\tau_1) + \psi \phi \omega (\overline{t-\tau_1}){{C}}_\mathrm{d}(t-\tau_1)-\sigma {{\hat{A}}}(t)
\end{equation}

The activation of T-cells against tumor cells is modeled through four bilinear equations which describe the generation of activated T-cells ({{$\hat{T}_\mathrm{a}$}}) or blocked T-cells ({{$\hat{T}_\mathrm{b}$}}) (through the CTLA-4 receptor) from a pool of blank T-cells ({{$\hat{T}$}}):
\begin{align}
	\label{eq_antigensRelease}
	\frac{d{{\hat{A}}}}{dt}(t)&=-a{{\hat{A}}}(t){{\hat{T}}}(t)-b{{\hat{A}}}(t){{\hat{T}}}(t)\\
	\label{eq_TcellActivation}
	\frac{d{{\hat{T}}}}{dt}(t)&=-a{{\hat{A}}}(t){{\hat{T}}}(t)-b{{\hat{A}}}(t){{\hat{T}}}(t)+h\\
	\label{eq_TcellActive}
	\frac{d{{\hat{T}_\mathrm{a}}}}{dt}(t)&=a{{\hat{A}}}(t){{\hat{T}}}(t)\\
	\label{eq_TcellInactive}
	\frac{d{{\hat{T}_\mathrm{b}}}}{dt}(t)&=b{{\hat{A}}}(t){{\hat{T}}}(t)
\end{align}
The constants $a$ and $b$ ($r=1+b/a$) describe the affinities for activation/inactivation, respectively~\cite{serre2016}. The pool of blank T-cells starts from $\hat{T}(0)=T_0$, which is assumed to be the carrying capacity of T-cells. It can be depleted due to activation/inactivation, and in that situation it can renew at constant rate (due to maturation of new T-cells), $h$. A constraint is imposed to avoid the T-cell compartment from exceeding the carrying capacity: $\hat{T}(t)\le T_0$. Active T-cells, {$\hat{T}_\mathrm{a}$}, migrate and infiltrate in the tumor (with a biological delay $\tau_2$, which phenomenologically models the time needed by active T-cells to act on tumor cells) where they become part of the compartment ${T_\mathrm{a}}$ (note the hat notation):
\begin{equation}
	\label{eq_TcellToL}
	\frac{d{{T_\mathrm{a}}}}{dt}(t) = \frac{d{{\hat{T}_\mathrm{a}}}}{dt}(t-\tau_2) = a {{\hat{A}}}(t-\tau_2) {{\hat{T}}}(t-\tau_2)
\end{equation}

We have tested the hypothesis that vascular damage at high radiation doses~\cite{song2015} may reduce the effectiveness of radioimmunotherapy by limiting the infiltration of T-cells in the tumor. Therefore, we include a dose and time dependent T-cell infiltrating parameter to account for vascular damage and recovery. Inspired by \cite{song2015} and \cite{rguezBarbeito2019}, we consider critical vascular damage for doses beyond $15$~Gy, and a progressive recovery of vascular function as,
\begin{equation}
	\label{eq_recovery}
	f(t) = \min\{0.05t, 1\}
\end{equation}
where the time post-irradiation, $t$, is measured in days. This term represents the fraction of active T-cells reaching the tumor, and multiplies (\ref{eq_TcellToL}).

\subsection{T-cell Mediated Tumor Cell Death}
Interaction between active T-cells and tumor cells results in the partial depletion of both. Following the work of de Pillis et al. \cite{depillis2005}, we model this interaction with a bilinear term for the compartment ${{T_\mathrm{a}}}$, in addition to an exponential natural elimination:
\begin{equation}
	\label{eq_LImmuneDeath}
	\frac{d{{T_\mathrm{a}}}}{dt}(t) = -\iota {{T_\mathrm{a}}}(t)C_\mathrm{tot}(t)-\eta {{T_\mathrm{a}}}(t)
\end{equation}
On the other hand, T-cell mediated tumor cell death is modeled with the following term \cite{depillis2005}:
\begin{equation}
	\label{eq_NImmuneDeath}
	\displaystyle \frac{d{{C}}}{dt}(t) = -p\frac{({{T_\mathrm{a}}}(t)/{{C_\mathrm{tot}}}(t))^q}{s+({{T_\mathrm{a}}}(t)/{{C_\mathrm{tot}}}(t))^q}{{C}}(t)
\end{equation}
The same expression holds for ${{C}}_\mathrm{d}(t)$. Note that these terms are coupled to the earlier equations.

\subsection{The Effect of $\alpha$PDL1 and $\alpha$CTLA4}
The concentration biokinetics (in arbitrary units) of $\alpha$PDL1 ($p_1$) and $\alpha$CTLA4 ($c_4$) is modeled as an instantaneous source term at injection times ($\{t_\mathrm{p1}\}$ and $\{t_\mathrm{c4}\}$, respectively) and a continuous exponential elimination:
\begin{align}
	\label{eq_c4}
	\frac{dc_4}{dt}(t)&=i_\mathrm{c_4}(t)\delta(t-\{t_\mathrm{c_4}\})-\nu c_4(t)\\
	\label{eq_p1}
	\frac{dp_1}{dt}(t)&=i_\mathrm{p_1}(t)\delta(t-\{t_\mathrm{p_1}\})-\mu p_1(t)
\end{align}

The precise pharmacokinetic modeling of these drugs is beyond the scope of the present article. However, this seems a good approximation for the kinetics of $\alpha$CTLA4 as Selby et al. \cite{selby2013} investigated different $\alpha$CTLA4 drugs biokinetics, finding that they follow linear forms like (\ref{eq_c4}). Although Deng et al. \cite{deng2016} investigated the biokinetics of $\alpha$PDL1, finding a more complex non-linear behaviour.

Rather than considering a complex kinetics model for the characterization of the effect of $\alpha$CTLA4 on the de-inhibition of T-cells, we model it as a simple dependence on the parameter $b$ in (\ref{eq_antigensRelease}), (\ref{eq_TcellActivation}) and (\ref{eq_TcellInactive}) on $c_4$, similarly to \cite{serre2016}:
\begin{equation}
	\label{eq_c4Effect}
	\displaystyle b \rightarrow \frac{b}{1+c_4(t)}
\end{equation}

On the other hand, the effect of $\alpha$PDL1 on the immune-death of tumor cells is modeled by introducing a dependence on the parameter $p$ in (\ref{eq_NImmuneDeath}) as:
\begin{equation}
	\label{eq_p1Effect}
	\displaystyle p \rightarrow p(1+p_1(t))
\end{equation}

\subsection{The Complete Model}

The assembled model is the following for ${{C}}$, ${{C}}_\mathrm{d}$, ${{\hat{A}}}$ and ${{T_\mathrm{a}}}$:
\begin{align}
	\label{eq_model}
	\displaystyle  \frac{d{{C}}}{dt}(t)&=\lambda_1 {{C}}(t)\left(1-\lambda_2{{C_\mathrm{tot}}}(t)\right)- K_{{\mathrm{C}}}(t)
		- p (1+p_1(t))\frac{({{T_\mathrm{a}}}(t)/{{C_\mathrm{tot}}}(t))^q}{s+({{T_\mathrm{a}}}(t)/{{C_\mathrm{tot}}}(t))^q}{{C}}(t)\nonumber\\
	\frac{d{{C}}_{\mathrm{d},i}}{dt}(t)&=K_{{\mathrm{C}}}(t) -\phi \omega(\bar{t}_i) {{C}}_{\mathrm{d},i}(t)
		 - p(1+p_1(t)) \frac{({{T_\mathrm{a}}}(t)/{{C_\mathrm{tot}}}(t))^q}{s+({{T_\mathrm{a}}}(t)/{{C_\mathrm{tot}}}(t))^q}{{C}}_{\mathrm{d},i}(t)\\
	\frac{d{{T_\mathrm{a}}}}{dt}(t)&=-K_{{\mathrm{T}}}(t) + a{{\hat{A}}}(t-\tau_2){{\hat{T}}}(t-\tau_2) 
		- \iota {{T_\mathrm{a}}}(t){{C_\mathrm{tot}}}(t) - \eta {{T_\mathrm{a}}}(t)\nonumber\\
	\frac{d{{\hat{A}}}}{dt}(t)&= \rho {{C_\mathrm{tot}}}(t-\tau_1)+ \psi \phi \sum_{i} \omega(\bar{t}_i-\tau_1){{C}}_{\mathrm{d},i}(t-\tau_1)
	-\sigma {{\hat{A}}}(t)- a{{\hat{A}}}(t){{\hat{T}}}(t)  - \frac{b}{1+c_4(t)}{{\hat{A}}}(t){{\hat{T}}}(t)\nonumber
\end{align}
In addition, (\ref{eq_TcellActivation})-(\ref{eq_TcellInactive}) control the activation of T-cells against tumor cells, and (\ref{eq_c4}, \ref{eq_p1}) control the biokinetics of $\alpha$PDL1 and $\alpha$CTLA4.

\subsection{Modeling Tumor Control Probability: Markov model}
\label{section:tcp}

We have employed the \textit{clonogenic cell hypothesis} \cite{nahum1993} to obtain tumor control probablities (TCP) from our model. It states that in order to control the tumor, all cells with proliferative capacity, which we identify with the compartment ${{C}}$, need to be eliminated. As defined in the previous section, the model is continuous and deterministic. In order to calculate TCPs we need a discrete model (numbers of cells) and stochasticity. Therefore, for low numbers of cells (${{C}}<1000$ cells), the model is converted to a Markov birth/death stochastic process \cite{hanin2001} by interpreting terms in the differential equations as birth/death probabilities. In a simulation, the tumor is considered controlled if ${{C}}$ reaches~0. 

In addition to the stochasticity of the Markov model, to obtain populational TCPs we also implemented random perturbations of the model parameters to simulate the heterogeneity of a population. For the population (a given number of simulations), TCP is computed as:
\begin{equation}
	\label{eq_TCP}
	\mathit{TCP}= \frac{\textit{number of controls}}{\textit{number of simulations}}
\end{equation}

More details about the TCP calculation and implementation are provided in the Supplementary Material.

\subsection{Experimental Data and Model Fitting}

In \cite{dewan2009} the authors studied the response of tumors in mice to RT (different fractionations) and $\alpha$CTLA4, either as monotherapies or in combination. Tumor cells (TSA breast carcinoma cells) were planted on the side of mice and let grow for 12~days, when they reached a volume of $\sim$32~mm$^3$. Treatments started at that time, and evolution of tumor volumes were monitored every 3~days. They studied tumor response to different combinations of RT+IT. In particular: {\bf i)} no treatment; {\bf ii)} RT alone, 20~Gy single-fraction; {\bf iii)}  RT alone, 3 fractions of 8~Gy (days 12, 13 and 14); {\bf iv)} RT alone, 5 fractions of 6~Gy (days 12, 13, 14, 15, and 16); {\bf v)}  IT alone, delivered in 3 fractions (days 14, 17, 20); {\bf vi)} combined RT+IT, 20~Gy + 3 fractions of IT (ii+v); {\bf vii)} combined RT+IT, (6 Gy$\times$5) + 3 fractions of IT (iv+v); {\bf viii)} combined RT+IT, (8 Gy$\times$3) + 3 fractions of IT (iii+v); {\bf ix)} combined RT+IT, (8 Gy$\times$3) + 3 fractions of IT at days 12, 15, 18; {\bf x)} combined RT+IT, (8~Gy$\times$3) + 3 fractions of IT at days 16, 18, 20. The study also reported the fraction of animals where tumor control was achieved (no evidence of tumor at the time of euthanasia).

In \cite{deng2014} the authors presented responses of tumors to radiotherapy and $\alpha$PDL1. Cancer cells (TUBO breast carcinoma cells) were implanted in mice, and tumors were allowed to grow for 14~days, when they had volumes around 120~mm$^3$. At that time, treatments started and tumor volumes were monitored up to day 35. The different experimental arms were: {\bf i)} no treatment; {\bf ii)} single dose of 12~Gy; {\bf iii)} four fractions of $\alpha$PDL1 at days 14, 17, 20 and 23; {\bf iv)} 12~Gy +  $\alpha$PDL1 (ii+iii).

In order to fit the reported evolution of (population-averaged) tumor volumes, we used the continuous model. Firstly, we let the modeled tumors to freely grow until they reached the relevant pre-treatment volumes reported in \cite{dewan2009} and \cite{deng2014}. That time was defined as reference (day~0), and treatment times are defined relative to it. Notice that the time to reach those volumes differs from the experimental results, as we start with different numbers of cells than those experimentally injected and our model does not aim to describe the process of tumor growth, which may be dominated by different mechanisms than tumor response.

Tumor volumes in our model are computed by considering the populations of both tumor cells (viable and doomed) and T-cells located in the tumor:
\begin{equation}
	\label{eq_volume}
	V_{\mathrm{model}} (t)= {{C_\mathrm{tot}}}(t)V_{{\mathrm{C}}}+{{T_\mathrm{a}}}(t)V_{{\mathrm{T}}}
\end{equation}
where $V_{{\mathrm{C}}}$ and $V_{{\mathrm{T}}}$ are the volumes of individual tumor cells and T-cells, respectively.

A simulated annealing method \cite{kirkpatrick1983} was implemented to find best fitting parameters. The objective function to be minimized is the weighted sum of square differences between model and experimental values:
\begin{equation}
	\label{eq_squareDiff}
	F = \sum_{\mathrm{curves}}\sum_{\mathrm{points}} \frac{(V_\mathrm{model}-V_\mathrm{exp})^2}{u^2}
\end{equation}
where $V_\mathrm{exp}$ and $V_\mathrm{model}$ are the experimental and model results, and $u$ are the experimental uncertainties. The optimization method has been applied to all response curves of each study at once (i.e. the sum above runs over different time points and different combinations of radiotherapy and immunotherapy), to avoid different best-fitting parameters for each curve.

{\subsection{Evaluation of goodness-of-fit}}

We have used the Akaike Information Criterion (AIC) with sample size correction \cite{banks2017} to compare best fits obtained with different direct damage terms (\ref{eq_LQ}), (\ref{eq_LQBetaMod}) and (\ref{eq_LQL}). This methodology ranks models according to the likelihood of the fit, $L$, and number of free parameters of the model, $k$:
\begin{equation}
\label{eq_AIC}
\mathit{AIC} = -2\log \left(L \right) + 2k + \frac{2k(k+1)}{N-k-1}
\end{equation}
where $N$ is the number of experimental data points, and $L$ is the maximum likelihood (of the best fit), calculated assuming a normal distribution of the experimental points and uncertainties. The model with the lowest AIC is considered the best model.

\subsection{Biologically Effective Dose}

The biologically effective dose (BED) \cite{jones2001} was used to design different radiobiologically iso-effective fractionations. The BED of a schedule delivering a total dose $D$ in fractions of dose $d$ is given by:
\begin{equation}
	\label{eq_BED}
	\mathit{BED} = D \left(1+\frac{d}{\alpha/\beta}\right)
\end{equation}
where $\alpha/\beta$ is the ratio of the linear and quadratic terms in the LQ model.

\subsection{Qualitative Sensitivity Analysis}

We have performed a qualitative local parametric sensitivity analysis. In order to do so, we have evaluated the sensitivity of the cost function to parameter perturbations around best-fitting values as:
\begin{equation}
\label{eq_sensitivity}
S_i = \left| F({\mathbf{x}}+\mathbf{\Delta x_i})- F({\bf x}) \right| 
\end{equation}
where $F$ is the cost function (\ref{eq_squareDiff}), ${\mathbf x}$ the set of best-fitting parameters and ${\bf \Delta x_i}=(0, ..., 0, 0.1x_i, 0, ..., 0)$ (setting a 10\% perturbation with respect to the best-fitting value $x_i$).

\begin{figure*}[t!]
	\centering
	\includegraphics[width=5.5in]{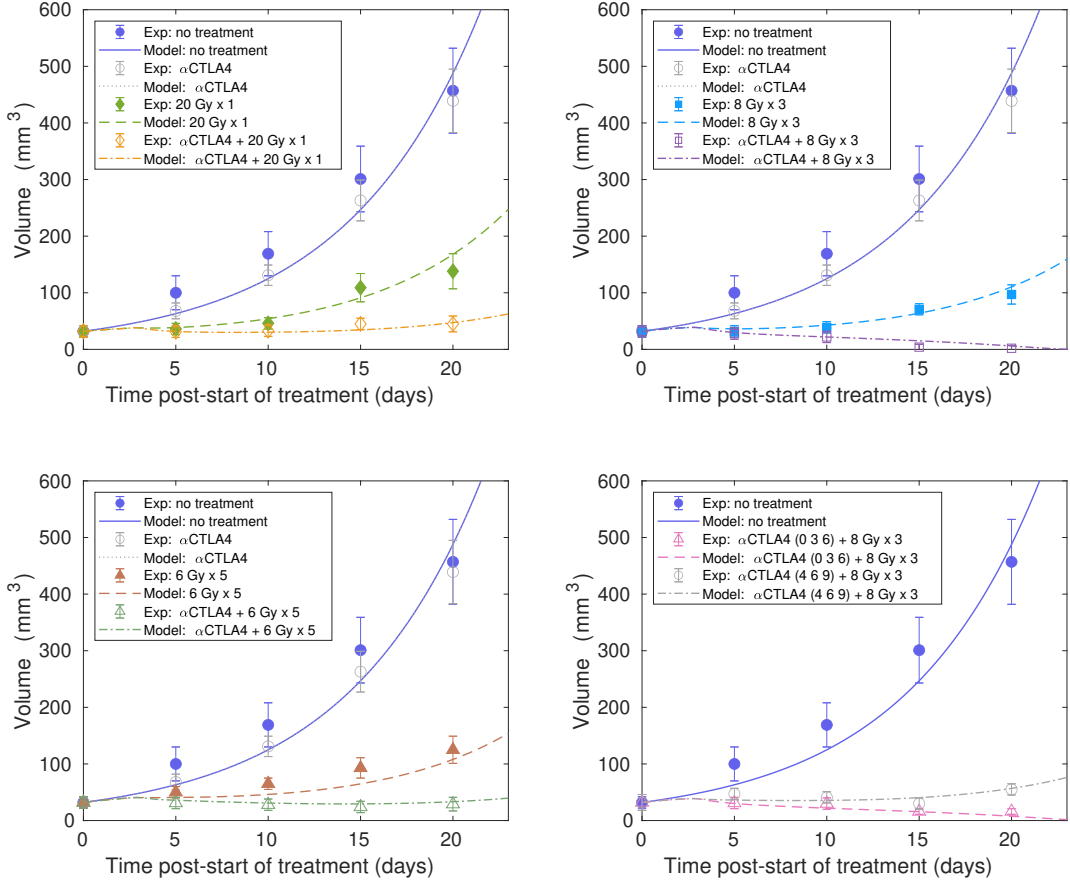}
	\caption{Model fitting of experimental data reported by Dewan et al. \cite{dewan2009} of tumor response to radiotherapy, immunotherapy with $\alpha$CTLA4, and combined treatment. Radiation doses are delivered in consecutive days starting from day 0, and immunotherapy doses are delivered at days (2, 5, 8) unless specified otherwise. Notice that differences between model curves for ``no treatment" and ``$\alpha$CTLA4" are small and both curves overlap in the figure. All curves have been obtained with a single set of parameters, although they are plotted separately to avoid overlaping and to make it easier to visualise the benefit of each combined treatment versus the independent use of RT or IT.}
	\label{fig_DewanFitting}
\end{figure*}

\subsection{Implementation and Parameters}
The model was implemented using different functions in Matlab (The Matworks, Natick, MA). The model is solved by employing an explicit Euler method \cite{press2007}, with a time step of $0.05$~days (details about the behavior of the method are available in the Supplementary Material). The main functions and the data used for model fitting are available from the Dataverse repository~\cite{glezCrespo2019}.

Not all model parameters were free during the fit to experimental data. Cell volumes in (\ref{eq_volume}) were set to $V_{{\mathrm{C}}} =10^{-6}$~mm$^3$ and $V_{{\mathrm{T}}} = 2\times 10^{-7}$~mm$^3$ \cite{kuse1985}. For the radiosensitivity of T-cells we have fixed the LQ parameters to $\alpha_T = 0.1$ and $\beta_{{\mathrm{T}}}=\alpha_{{\mathrm{T}}}/10$. The biokinetic elimination rates of $\alpha$PDL1 and $\alpha$CTLA4 were set to $\mu=0.5$~days$^{-1}$ and $\nu=0.1$~days$^{-1}$ respectively, from fits to data reported in \cite{selby2013, deng2016}. The parameters characterizing the mitotic delay of radiation-damaged cells were set to {${\tau_\mathrm{d}}_1 =1$~days, ${\tau_\mathrm{d}}_2 =1.5$~days}, which is in the range of reported mitotic delays. The parameter $r$, relative to the activation/inactivation rate of T-cells is set to $5$, as in \cite{serre2016}.

On the other hand, values of best-fitting parameters were constrained to qualitative reasonable bounding intervals when deemed necessary, in order to avoid unphysical/unreasonable values (for example, $\alpha>$ 0 and $\beta>$ 0 in the LQ-model, or $\lambda>$ 0 for tumor proliferation). The bounding intervals are shown in Supplementary~Table~4.

\begin{figure*}[t!]
	\centering
	\includegraphics[width=5.5in]{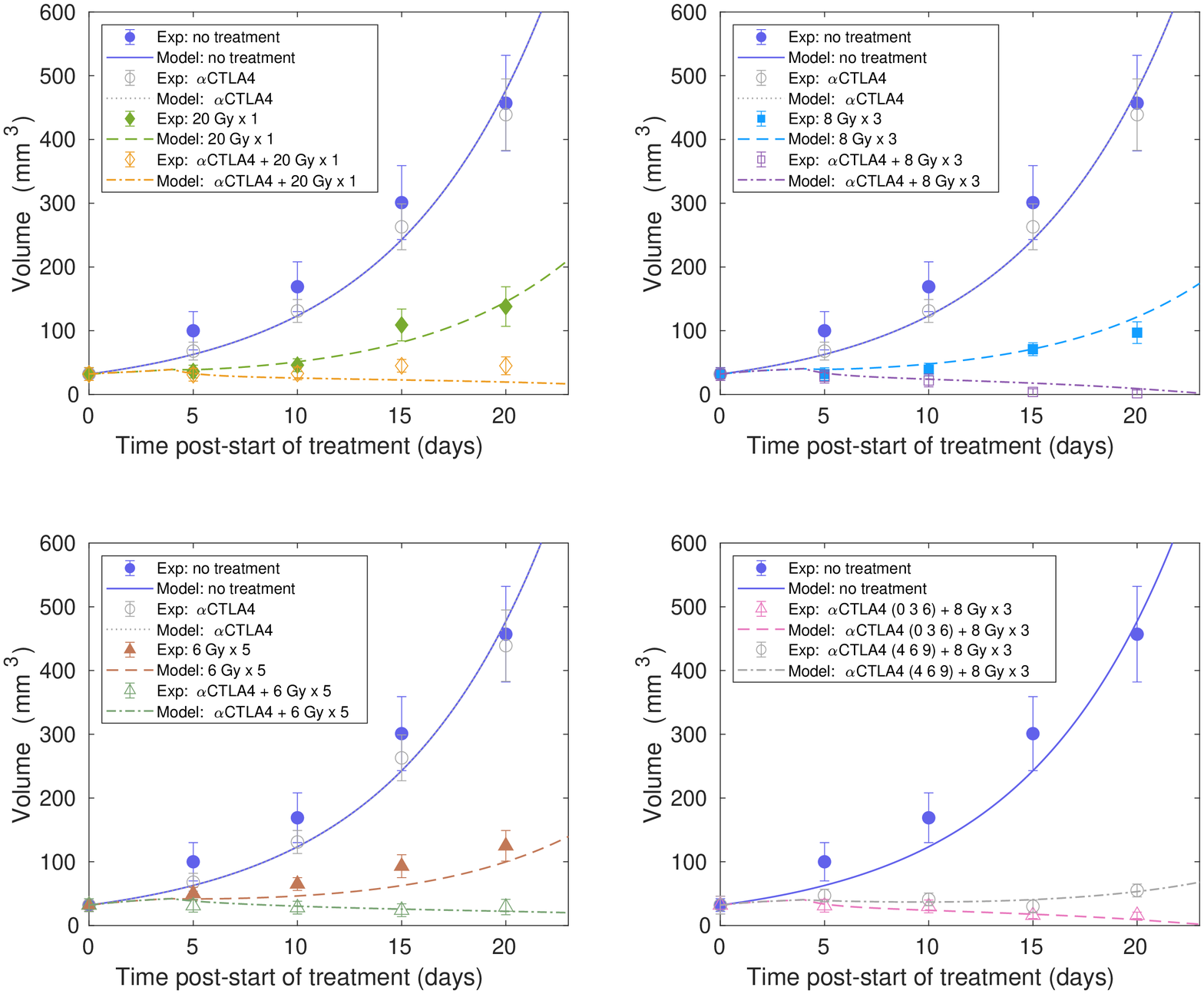}
	\caption{Model fitting of experimental data reported by Dewan et al. \cite{dewan2009} of tumor response to radiotherapy, immunotherapy with $\alpha$CTLA4, and combined treatment. The LQL model was used to account for the radiosensitivity of tumor cells. Radiation doses are delivered in consecutive days starting from 0, and immunotherapy doses are delivered at days (2, 5, 8) unless specified otherwise. Notice that differences between model curves for ``no treatment" and ``$\alpha$CTLA4" are small and both curves overlap in the figure. All curves have been obtained with a single set of parameters, although they are plotted separately to avoid overlaping and to make it easier to visualise the benefit of each combined treatment versus the independent use of RT or IT.}
	\label{fig_DewanLQL}
\end{figure*}

\begin{figure}[t!]
	\centering
	\includegraphics[width=3in]{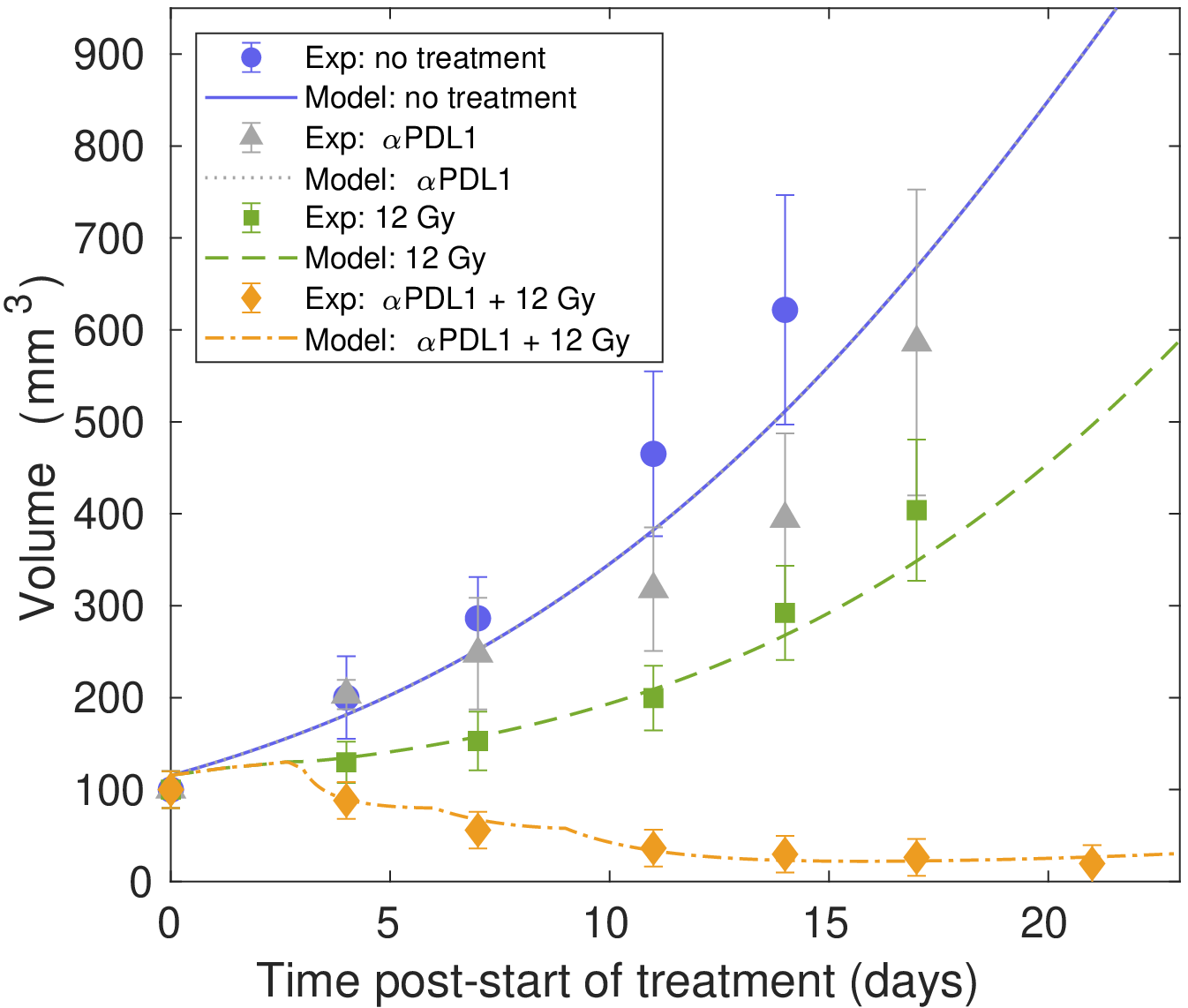}
	\caption{Model fitting of experimental data reported by Deng et al. \cite{deng2014} of tumor response to radiotherapy (12~Gy$\times$1 fraction), immunotherapy with $\alpha$PDL1, and combined treatment. Notice that differences between model curves for ``no treatment" and ``$\alpha$PDL1" are small and both curves almost overlap in the figure.}
	\label{fig_DengFitting}
\end{figure}

\section{Results}

\subsection{The Model Can Fit Pre-clinical Data of Tumor Response to Combined Therapies of Radiation and $\alpha$PDL1/$\alpha$CTLA4}
\label{section_PDL1}

In Fig.~\ref{fig_DewanFitting} we report best fits of our model to volume dynamics data presented in \cite{dewan2009}, when employing the modified LQ-model for direct radiation death (\ref{eq_LQBetaMod}). In this dataset, when no treatment is delivered, tumor volumes grow exponentially, $\alpha$CTLA4 alone has no significant effect on tumor response, RT alone causes a moderate tumor response, and the combination of RT and $\alpha$CTLA4 leads to an important tumor response, achieving tumor control in some cases (mostly for 8~Gy$\times$3). Our model reproduce these progression patterns, as shown in Fig.~\ref{fig_DewanFitting}. Best-fitting parameters are presented in Supplementary Table 1. We highlight the most relevant parameters associated with proliferation ($\lambda_1 \simeq 0.14~\mathrm{day}^{-1}$), radiation damage ($\alpha_{{\mathrm{C}}} \simeq 0.02~\mathrm{Gy}^{-1}$, $\beta_{{\mathrm{C}}}\simeq0.007~\mathrm{Gy}^{-2}$, $c\simeq-0.2~\mathrm{Gy}^{-1/2}$) and the immune effect on tumor cells ($p \simeq 24.4~\mathrm{day}^{-1}$) and T-cells ($\iota \simeq 2\times10^{-8}~\mathrm{day}^{-1}$).

Data fitting point to a decrease in relative radiosensitivity with increasing dose (negative parameter $c$ in (\ref{eq_LQBetaMod})). Such behavior can be described by the LQL model. Therefore, we have performed the same fit with the LQL model instead (Fig. \ref{fig_DewanLQL}), obtaining similar results. Best-fitting parameters are reported in Supplementary Table 2, including the most relevant parameters associated with proliferation ($\lambda_1 \simeq 0.14~\mathrm{day}^{-1}$), radiation damage ($\alpha_{{\mathrm{C}}} \simeq 0.04~\mathrm{Gy}^{-1}$, $\beta_{{\mathrm{C}}}\simeq0.017~\mathrm{Gy}^{-2}$, $x\simeq8.4~\mathrm{Gy}^{-1}$) and the immune effect on tumor cells ($p \simeq 23.3~\mathrm{day}^{-1}$) and T-cells ($\iota \simeq 2\times10^{-8}~\mathrm{day}^{-1}$). The best-fitting value of the cost function is $F=29.56$ and $F=42.18$ for the modified LQ and LQL model, respectively.

In Fig.~\ref{fig_DengFitting} we report best fits of our model to data presented in \cite{deng2014}, which shows tumor responses to radiotherapy and $\alpha$PDL1. This dataset presents similar patterns of tumor response: $\alpha$PDL1 alone has not significant effect on tumor response, RT alone causes a moderate tumor response, and the combination of RT+$\alpha$PDL1 presents synergy and leads to an important tumor response.

\begin{figure*}[!t]
	\centering
	\includegraphics[width=5.5in]{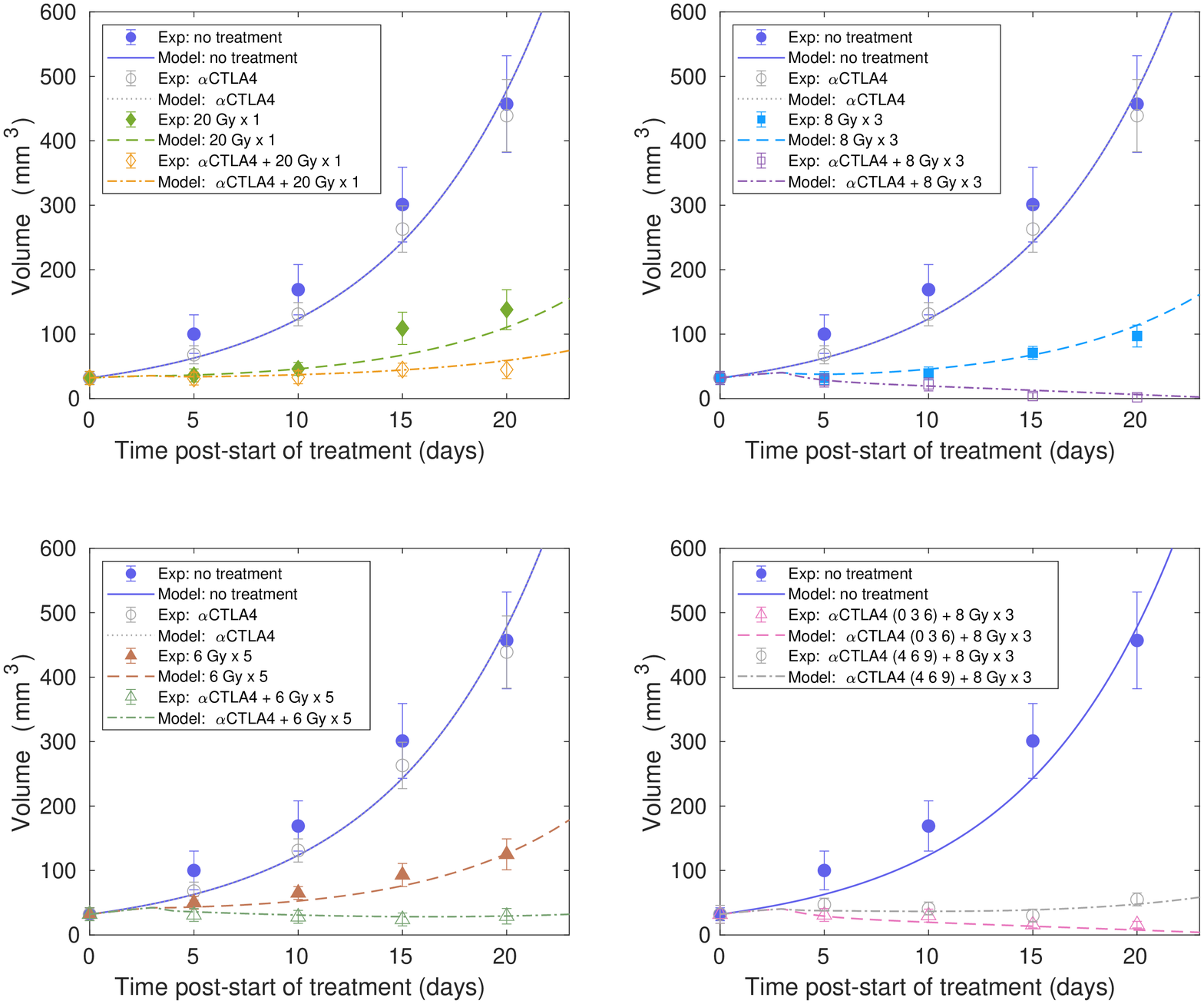}
	\caption{Model fitting of experimental data reported by Dewan et al. \cite{dewan2009} of tumor response to radiotherapy, immunotherapy with $\alpha$CTLA4, and combined treatment. The classical LQ model is used to characterize tumor cell response to radiation, and vascular damage at 20~Gy per fraction is included to limit T-cell infiltration in the tumor. Radiation doses are delivered in consecutive days starting from 0, and immunotherapy doses are delivered at days (2, 5, 8) unless specified otherwise (bottom-right panel). Notice that differences between model curves for ``No treatment" and ``$\alpha$CTLA4" are small and both curves overlap in the figure. All curves have been obtained with a single set of parameters, although they are plotted separately to avoid overlaping and to make it easier to visualise the benefit of each combined treatment versus the independent use of RT or IT.}
	\label{fig_DewanFitting2}
\end{figure*}

To fit the data of Fig.~\ref{fig_DengFitting} we have kept fixed most of the best-fitting parameters obtained when fitting Fig.~\ref{fig_DewanFitting}: only parameters related to the dose of $\alpha$PDL1, tumor cell proliferation ($\lambda_1\simeq0.12~\mathrm{day}^{-1}$) and tumor cell radiosensitivity ($\alpha_C \simeq 0.03~\mathrm{Gy}^{-1}$) were allowed to vary. Because this dataset only includes one dose per fraction, we have used the LQ model with $\alpha_{\rm{C}}/\beta_{\rm{C}}=10$ Gy (i.e. only $\alpha_{\rm{C}}$ is a free parameter). While there are differences in the clones and tumors that could justify using different host-related and tumor-related parameters, we think that imposing such constraints on the optimization poses a serious test to our model, and avoids reaching good fits by over-fitting. Best-fitting parameters are reported in Supplementary Table 1.

\subsection{Vascular Damage May Limit the Effectiveness of Radioimmunotherapy}

Large radiation doses can seriously damage tumor vasculature, which might limit the infiltration of active T-cells in the tumor. This might also explain the poorer results obtained with the 20~Gy single-fraction irradiation in \cite{dewan2009}. In order to test the hypothesis that vascular damage may affect the effectiveness of radioimmunotherapy, we removed the dependence of the tumor cells $\beta$-term on the radiation dose (see Section~\ref{section:lq}), and we included a dose and time dependent T-cell infiltrating parameter in our model (\ref{eq_recovery}), to account for vascular damage and recovery. Inspired by \cite{song2015, rguezBarbeito2019}, we consider critical vascular damage for irradiation above 15~Gy, followed by a progressive recovery of vascular function. This factor represents the fraction of active T-cells reaching the tumor.

In Fig.~\ref{fig_DewanFitting2} we show best fits of our model to data presented in \cite{dewan2009}. The model provides a good fit to the experimental values. The goodness of the fit is slightly better than those reported in Fig.~\ref{fig_DewanFitting} and Fig.~\ref{fig_DewanLQL}: $\mathit{AIC}=436.24$ ($k=19$, $F=25.33$, $N=50$), versus $\mathit{AIC}=446.02$ ($k=20$, $F=29.59$) and $\mathit{AIC}=458.75$ ($k=20$, $F=42.18$) obtained with the modified LQ and LQL models, respectively. Best-fitting model parameters are shown in Supplementary Table~3, including the most relevant parameters associated with proliferation ($\lambda_1 \simeq 0.14~\mathrm{day}^{-1}$), radiation damage ($\alpha_{{\mathrm{C}}} \simeq 0.02~\mathrm{Gy}^{-1}$, $\beta_{{\mathrm{C}}}\simeq0.002~\mathrm{Gy}^{-2}$) and the immune effect on tumor cells ($p \simeq 24.9~\mathrm{day}^{-1}$) and T-cells ($\iota \simeq 6\times10^{-9}~\mathrm{day}^{-1}$).

\begin{figure*}[t!]
	\centering
	\includegraphics[width=5.5in]{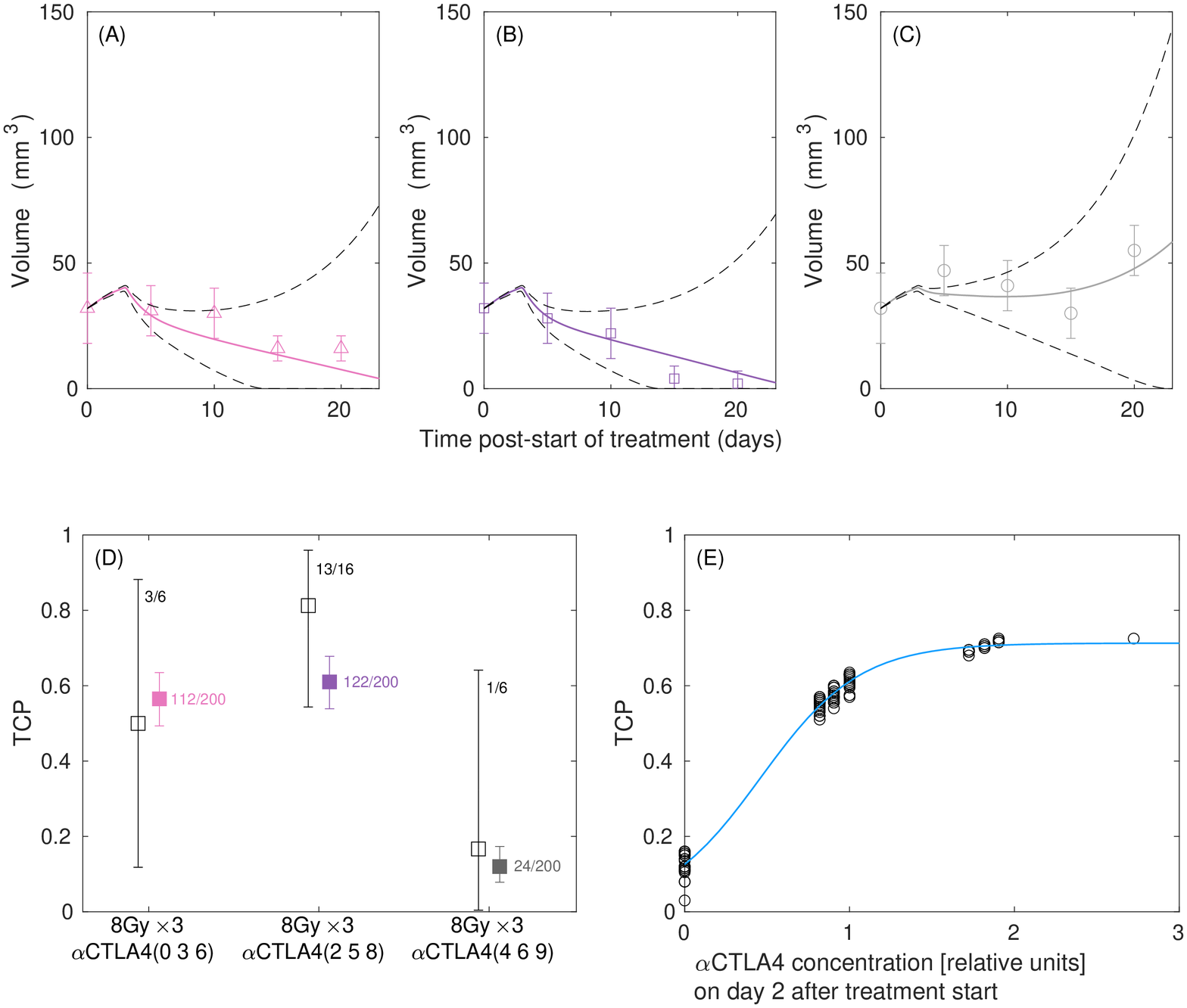}
	\caption{Study of optimal schedules of administration of radiotherapy (8~Gy$\times$3 fractions) and immunotherapy (3 fractions of $\alpha$CTLA4) obtained with the biomathematical model and parameters reported in Supplementary Table 3. Radiotherapy fractions are delivered at days (0, 1, 2), and immunotherapy is delivered with different schedules starting from day 0 to day 7. In panels (A), (B) and (C) we report the dynamics of tumor volumes and the 95\% confidence intervals for three combinations that were investigated in \cite{dewan2009}, delivering immunotherapy at days (0, 3, 6), (2, 5, 8), and (4, 6, 9) respectively. In panel (D) we compare 95\% confidence intervals for TCP values obtained with the model (200 simulations) and experimental controls (6 to 16 animals), for the same three treatment combinations. In (E) we present model TCPs (circles) versus the concentration of $\alpha$CTLA4 on day~2 after starting treatment, showing that there is a positive correlation between those two variables. The blue solid line corresponds to the fit to a logistic function.}
	\label{fig_schedules}
\end{figure*}

In Table \ref{tab_sensitivity_vd} we rank the sensitivity of the cost function to model parameters. The model is most sensitive to parameters describing tumor cell proliferation, radiosensitivity and immune-mediated tumor cell killing.

\begin{table*}[!b]
	\renewcommand{\arraystretch}{1.3}
	\caption{Analysis of the most critical model parameters. The sensitivity index given by (\ref{eq_sensitivity}) for each parameter is calculated as the difference between the cost function of fits to experimental data presented in \cite{dewan2009} (best-fitting parameters reported in Supplementary Table 3) and the cost function obtained when a 10\% perturbation is applied to that particular parameter. Only the ten most critical parameters are shown. For reference, the cost function value obtained with best fitting parameters is $25.33$.}
	\label{tab_sensitivity_vd}
	\centering
	\begin{tabular}{|c|c|c|}
		\hline		
		\bfseries Parameter & \bfseries Description [units] & \bfseries Sensitivity index\\
		\hline\hline
		$q$ & Immune-mediated tumor cell death parameter & $309.4673$\\
		\hline
		$\lambda_1$ & Proliferation rate of tumor cells [day$^{-1}$] & $71.6561$\\
		\hline
		$s$ & Slope of the immune-mediated tumor death curve & $20.7941$\\
		\hline
		$p$ & Immune-mediated tumor cell death rate [day$^{-1}$] & $12.8123$\\
		\hline
		${{\hat{T}}} (t=0)$ & Initial pool of \textit{blank} T-cells in activation site & $7.3442$\\
		\hline
		$\eta$ & Rate of natural elimination of T-cells [day$^{-1}$] & $4.21149$\\
		\hline
		$h$ & Rate of production of \textit{blank} T-cells {[day$^{-1}$]} & $3.5890$\\
		\hline
		$\alpha_{{\mathrm{C}}}$ & Linear parameter of LQ model for tumor cells [Gy$^{-1}$] & $2.1120$\\
		\hline
		${\beta_{{\mathrm{C}}}}$ & Quadratic parameter of LQ model for tumor cells [Gy$^{-1}$] & $1.8827$\\
		\hline
		$\tau_1$ & Delay between antigen liberation and T-cell activation [day] & $0.6541$\\
		\hline	
	\end{tabular}
\end{table*}

Volumes presented in previous figures include tumor cells and T-cells. Certainly, we do not want the model to reproduce tumor volumes by including low fractions of tumor cells and large fractions of T-cells, which would eventually lead to tumor control, and would contradict experimental evidence. In Supplementary~Fig.~1 and Supplementary~Fig.~2 we present the contribution of tumor cells and T-cells to tumor volumes, showing that modeled tumor volumes are dominated by tumor cells.

\subsection{Optimizing the Schedule of Administration of Radiotherapy and Immunotherapy Can Improve Effectiveness}

Finding the optimal sequence of administration of IT+RT can improve effectiveness, as experimentally shown in results presented in \cite{dewan2009}: in that study, the authors found that the combination of 8~Gy$\times$3 radiotherapy fractions and 3 fractions of $\alpha$CTLA4 lead to different tumor response depending on the days of administration of the immunotherapy (radiotherapy was always delivered on the same days); in particular, different immunotherapy schedules lead to different evolutions of tumor volumes, and to substantially different tumor control rates. This may be caused by the interplay between biological mechanisms with different kinetics, including the biological delays arising from the release of antigens to the activation of T-cells and the migration of such T-cells to the tumor, tumor proliferation and the progressive death of doomed cells.

We have investigated this effect with our model. We have used the best-fitting parameters reported in Supplementary Table 3. In order to simulate population tumor control probabilities, we have used the stochastic Markov model described in Section \ref{section:tcp}, together with a relative normal random dipersion of best-fitting parameters of 5\% (to simulate population heterogeneity). The days of administration of radiotherapy were kept fixed (0, 1 and 2 days), and the days of administration of IT were varied, starting from day~0 to day 7 and ending from day 2 to day 9. The evolution of tumor volumes and the TCP was evaluated for each treatment configuration from 200 simulations.

Results are reported in Fig. \ref{fig_schedules}. In Fig. \ref{fig_schedules}(A)-(C) we report the evolution of tumor volumes with best-fitting parameters and 95\% confidence intervals (obtained from the 200 different simulations) for different combinations that were investigated in~\cite{dewan2009}. In Fig. \ref{fig_schedules}(D) we compare 95\% confidence intervals for TCP values obtained with the model (200 simulations) and experimental controls (6 to 16 animals), for the same three treatment combinations. Of the three experimental combinations, the one causing better tumor response is that delivering immunotherapy at days 2, 5 and~8.

We have qualitatively investigated the correlation between modeled TCP and different metrics related to the administration of immunotherapy. We found that the concentration of $\alpha$CTLA-4  on day 2 shows a positive correlation with TCP. This is shown in Fig. \ref{fig_schedules}(E).

\begin{figure*}[t!]
	\centering
	\includegraphics[width=5.5in]{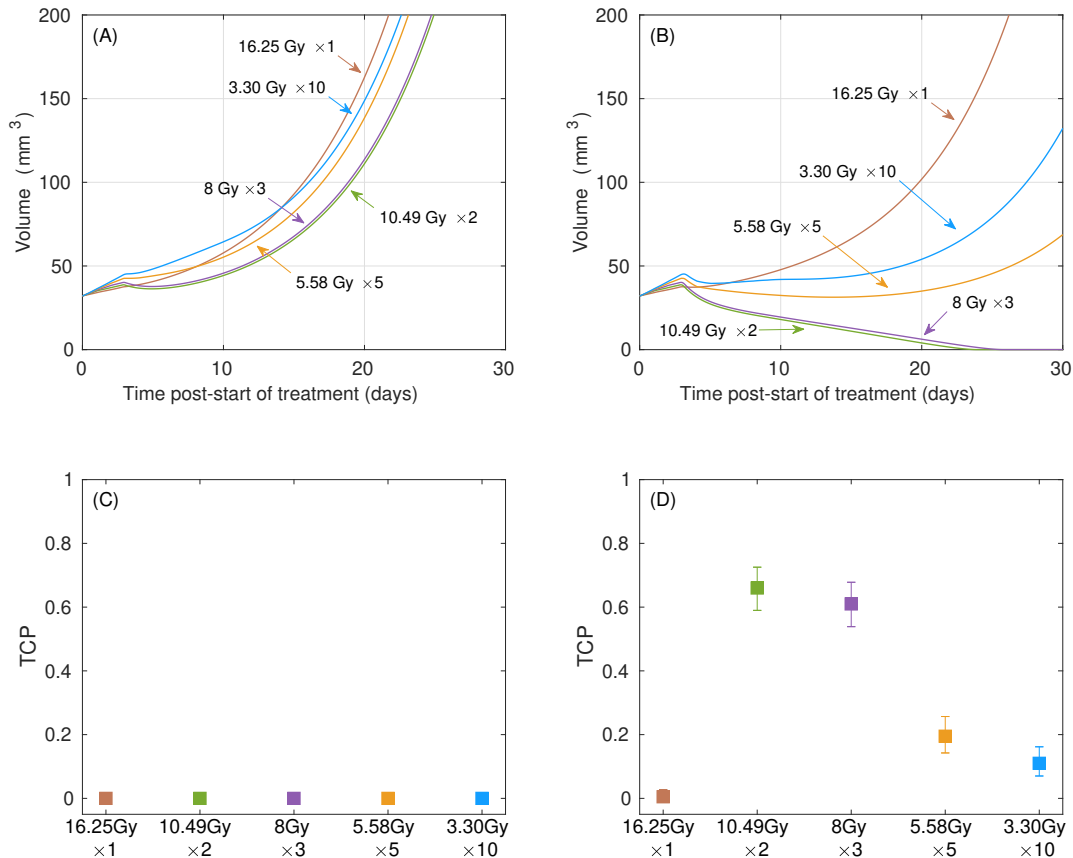}
	\caption{Modeled responses to different radiotherapy schedules combined (or not) with three fractions of $\alpha$CTLA4 (at days 2, 5, and 8). Radiation schedules are iso-BED, and therefore equivalent from a classical radiobiological point of view, yet they lead to very different response curves due to the different interaction between radiation-induced cell death and immune-induced cell death. Tumor volume evolution is shown when treating with radiotherapy alone (A) and radiotherapy plus $\alpha$CTLA4 {(B)}. We also report tumor control probabilities (95\% confidence intervals) obtained with the model (200 simulations) when simulating radiotherapy alone (C) and radiotherapy plus $\alpha$CTLA4 (D). Model parameters used for this study are presented in Supplementary Table 3, and include vascular damage effect at 16.25~Gy (\ref{eq_recovery}).}
	\label{fig_fractionation}
\end{figure*}

\subsection{Optimal Dose Fractionation Can Improve Effectiveness Of Radioimmunotherapy}

With the set of parameters shown in Supplementary Table~3 we have studied the combined effect of different fractionations of RT with or without three fractions of $\alpha$CTLA4. The 8~Gy$\times$3 fractions treatment was used as the reference treatment, and different radiation dose fractionations (1, 2, 3, 5 or 10 fractions, one fraction per day, consecutive days) were studied, without IT, and in combination with three fractions of $\alpha$CTLA4 at days 2, 5 and 8 post-start of treatment. The doses of each fractionation schedule were selected to have the same BED, calculated with $\alpha_{{\mathrm{C}}}/\beta_{{\mathrm{C}}}=9.29$~Gy (Supplementary Table 3). With this $\alpha_{{\mathrm{C}}}/\beta_{{\mathrm{C}}}$ value, the doses per fraction for each fractionation were (for $n$ fractions): $16.25$~(1); $10.49$ (2); $8$ (3); $5.58$ (5); $3.30$~(10).

Results are presented in Fig.~\ref{fig_fractionation}, where we report evolution of tumor volumes under different treatment regimens, as well as tumor control probabilities (obtained as in the previous section, from 200 simulations with the stochastic Markov model described in Section \ref{section:tcp}, together with a relative normal random dispersion of best-fitting parameters of 5\% to simulate population heterogeneity).

Conventional fractionation may be sub-standard, as daily fractions can deplete active T-cells from the tumor (it is widely assumed that T-cells are radiosensitive, even though this idea may be contradicted by recent evidence~\cite{arina2019}). In this regard, hypofractionated schedules may prove more effective (as already seen in some experimental studies), but without reaching extreme hypofractionation. In the latter case, the strategy may prove disadvantageous due to two factors: on the one hand, a single fraction may fail to keep therapeutic numbers of T-cells in the tumor for long times; on the other hand, very large doses may lose some effectiveness, and can seriously damage tumor vasculature, which might limit the infiltration of active T-cells in the tumor.

\begin{figure}[!h]
	\centering
	\includegraphics[width=4.4in]{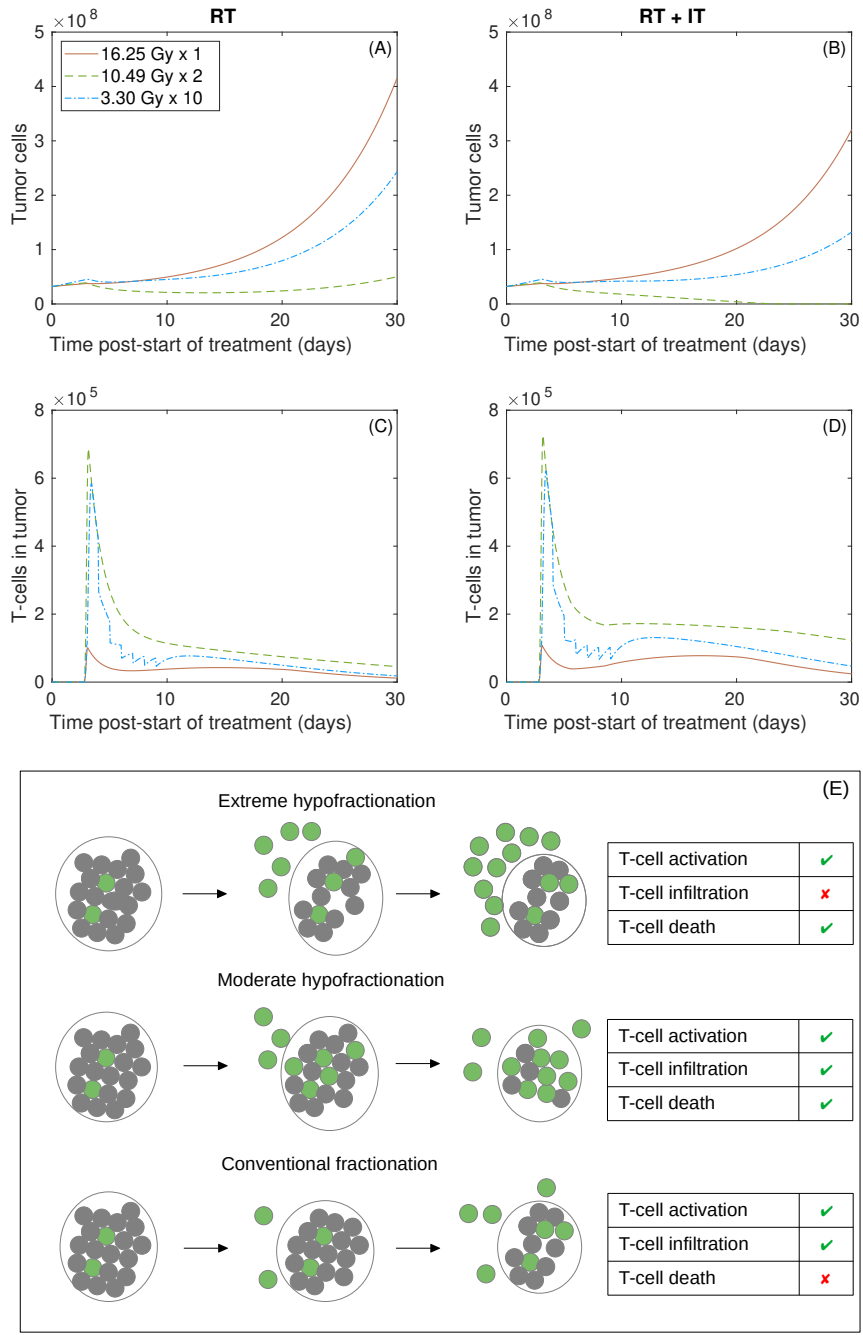}
	\caption{Comparison of tumor response to different types of fractionation: single-dose hypofractionation, non-single-dose hypofractionation and more fractionated treatment. The panels (A)-(D) show the evolution of tumour cells and active T-cells in tumor zone for the different fractionations. The panels on the left correspond to radiotherapy and those on the right to radiotherapy in combination with immunotherapy. A summary sketch illustrating the response to different fractionations is shown in the panel (E). Green circles represent activated T-cells, gray circles are tumor cells, and arrows represent the pass of time, starting from the time of treatment. Non-single-dose hypofractionated treatments appear to be the most effective, as they allow T-cell infiltration into the tumor and reduces T-cell damage, unlike the other two types of treatment.}
	\label{fig_conclusion}
\end{figure}

Fig.~\ref{fig_conclusion} shows the evolution of the main populations for three different treatments. As can be seen in panels (A) and (B), in line with what was hypothesised above, the combined treatment with hypofractionated non-single-dose radiation is the most effective. Furthermore, in panels (C) and (D) it can be seen that while with single-dose treatment the number of T-cells reaching the tumor is limited, with a more fractionated treatment the radiation eliminates active T-cells in tumor reducing its efficacy. Fig.~\ref{fig_conclusion}(E) summarizes the findings and shows the strengths and weaknesses of each treatment.

\section{Discussion}

Radioimmunotherapy shows potential to improve cure rates in many cancer types. \textit{In silico} mathematical models of tumor response to radioimmunotherapy can help to interpret experimental data, and ultimately may help to identify optimal therapeutic strategies. In this work we present a biomathematical model of tumor response to radioimmunotherapy with inhibitors of PD-L1 ($\alpha$PDL1) and CTLA-4 ($\alpha$CTLA4), which consists of a system of coupled impulsive delay differential equations to include biological response delays. Our approach follows an intermediate path between systems biology and phenomenological modeling, including a detailed description of the processes involved in response to radioimmunotherapy but without reaching the cellular/molecular scale of systems biology. We present the model in two flavors: on the one hand, a continuous/deterministic model, which is mainly used to fit experimental data of populational dynamics of tumor volumes; on the other hand, a discrete/stochastic model, obtained by treating the model as a Markov death/birth process, which is used to model tumor control probabilities according to the clonogenic cell hypothesis.

The model can fit experimental tumor responses (evolution of tumor volumes) to different combinations of radiotherapy and immunotherapy with $\alpha$PDL1 and $\alpha$CTLA4. It can also qualitatively reproduce experimental TCP values, even though the uncertainties of those values are large because they are obtained from a limited number of animals. A limitation of our model/analysis is the risk of overfitting, since we do not have a large number of experimental data to fit it compared to the number of parameters involved. To mitigate this weakness, we have set as many parameters as possible according to the existing literature and limited their range of variation in the optimization to consistent values. Furthermore, most of the parameters (16/20) obtained for the data fit with administration of $\alpha$CTLA-4 have been kept fixed in the fit with $\alpha$PD-L1. These constraints strengthen the results obtained as mentioned in Section~\ref{section_PDL1}, but they should still be taken with care due to the the risk of overfitting.

One thing our model fails to reproduce is the effect of $\alpha$PDL1 and $\alpha$CTLA4 as monotherapy on progression of tumor volumes. While this effect is small in the collected experimental data (differences observed in a small number of animals are not significant), they show a consistent effect of IT alone on tumor volume. Our modeling results show no noticeable effect of $\alpha$CTLA4 or $\alpha$PDL1 monotherapies. Therefore, the application of our model to IT monotherapies could be debatable. However, it might well be that the low percentage of subjects responding to IT monotherapy treatments present a particular phenotype (model parameters) that makes them more sensitive to such therapies, and such effect cannot be reproduced when fitting population-averaged data. 

The fits show very radio-resistant tumors (low $\alpha$ values), but in line with reported experimental values for tumors in mice \cite{vanLeeuwen2018}. Interestingly, best fits are obtained when the relative radiosensitivity of the tumor cells decreases with increasing dose. This has been observed by fitting the data to a modified LQ-model that includes such behaviour, to the LQL model (which accounts for decreasing radiosensitivity through increasing repair), and also by limiting the infiltration of T-cells into the tumor due to vascular damage at large doses. All these models lead to similar fits. We must notice that we have only included the effect of limited T-cell infiltration in this analysis for the sake of simplicity, but vascular damage may cause other effects, like proliferation arrest, starvation and hypoxia, resulting in a complex interaction.

Based on the results obtained, we can formulate hypotheses regarding the effectiveness of radioimmunotherapy and optimal treatment combinations. In particular:

\begin{itemize}
	\item
	We have investigated the optimal schedule of administration of RT (8 Gy $\times$ 3) and IT (3 fractions). Experimental results show that delivering immunotherapy at days (2, 5, 8) leads to a better response than deliveries at days (0, 3, 6), and also (4, 6, 9). This is particularly interesting given the slow elimination rate of $\alpha$CTLA4 ($T_{1/2}\sim7$ days, the elimination from day 0 to day 2, and from day 2 to day 4, is $\sim 20\%$ of the dose), and seems to point out that some biological delays condition the activation and/or immune effect of T-cells. A modeling study of optimal combinations suggests that a better synergy could be obtained by delivering IT soon after the first fraction of RT, and then continuing IT fractions longer into the treatment, to keep therapeutic numbers of T-cells in the tumor for a longer time (according to the different kinetics of the IT drug, tumor cell death, and antigen liberation). The combined treatment starts to lose effectiveness if the administration of IT is delayed beyond 2/3 days post-start of RT.

	\item We have also investigated optimal dose fractionation. Conventional fractionation of dose may be suboptimal for radioimmunotherapy, but extreme hypofractionation may also prove suboptimal, either due to lose of effectiveness of such high-doses (modeled in the work with the LQL model and a modified LQ model), or to critical damage to tumor vasculature that may limit T-cell infiltration. This might also explain the poorer results obtained with the 20~Gy single-fraction irradiation. Moderate hypofractionation of the radiation dose may offer the best results.
	
\end{itemize}

Direct extrapolation of these hypotheses to clinical data must be considered with care, for they are based on the analysis of preclinical data. Human tumors may behave differently: for example, many processes (proliferation, clearance, biokinetics, migration...) are slower in humans, which will certainly affect the complex interplay that leads to optimal treatment combinations.

The combined study of what is usually called indirect tumor cell death in radiotherapy/ radioimmunotherapy (due to vascular damage and radiation-triggered immune response) certainly seems to be of high importance, as these effects may interfere with each other. Biomathematical models linking them can provide insight into the problem and help to interpret experimental results. Biomathematical modeling may also help to interpret conflicting experimental reports on the effect of indirect cell death on tumor response (for example, \cite{song2015, moding2015}).

Compared to other models of immunotherapy/radioimmunotherapy response, in this work we present a simpler modeling of the immune-mediated tumor cell death, including a single population of T-cells and ignoring other populations. On the other hand, the most relevant novelty of our work is the introduction of biological delays (and the use of delay differential equations). We hypothesize that these biological delays play an important role in the response to radioimmunotherapy, and should be considered in the design of optimal combination strategies of radiotherapy and immunotherapy. This seems to be supported by experimental results, as argued above. We also present a stochastic approach to compute tumor control probabilities, and investigate the effect of vascular damage in the response to radioimmunotherapy.

\section{Conclusions}

Biomathematical models that can help to interpret experimental data of the synergies between radiotherapy and immunotherapy, and to assist in the design of more effective radioimmunotherapies, could potentially facilitate the implementation of optimized radioimmunotherapy. Such models may help to interpret and analyze clinical and pre-clinical data. From our work, it seems that a good understanding of the biological delays associated with such therapies, the biokinetics of the immunotherapy drug, and the interplay among them, may be of paramount importance for designing optimal radioimmunotherapy schedules. Ultimately, we envision a key role for these models, assisting in the design of more effective radioimmunotherapies, similarly to the role of the LQ model in conventional radiotherapy. Further work is required to create well-validated models, but studies like the one presented in this paper constitute a stepping stone towards that final goal.

\section*{Acknowledgments}
We acknowledge the support of Instituto de Salud Carlos III through research grant PI17/01428 (FEDER co-fund). O.L.P. was co-financed by the European Regional Development Fund (ERDF) and the Xunta de Galicia under the 
GRC2013-014 grant, and by the Spanish Ministry of Science, 
Innovation and Universities under the MTM2017-86459-R grant. We are grateful to Juan Vi\~{n}uela and Lois Okereke for useful discussions.

\bibliographystyle{unsrt}

\newpage
\renewcommand\figurename{Supplementary Figure}
\renewcommand\tablename{Supplementary Table}
\setcounter{figure}{0}
\setcounter{table}{0}
\setcounter{section}{0}

\section{Supplementary Figures}
\begin{figure*}[h!]
	\renewcommand{\arraystretch}{1.3}
	\centering
	\includegraphics[width=6in]{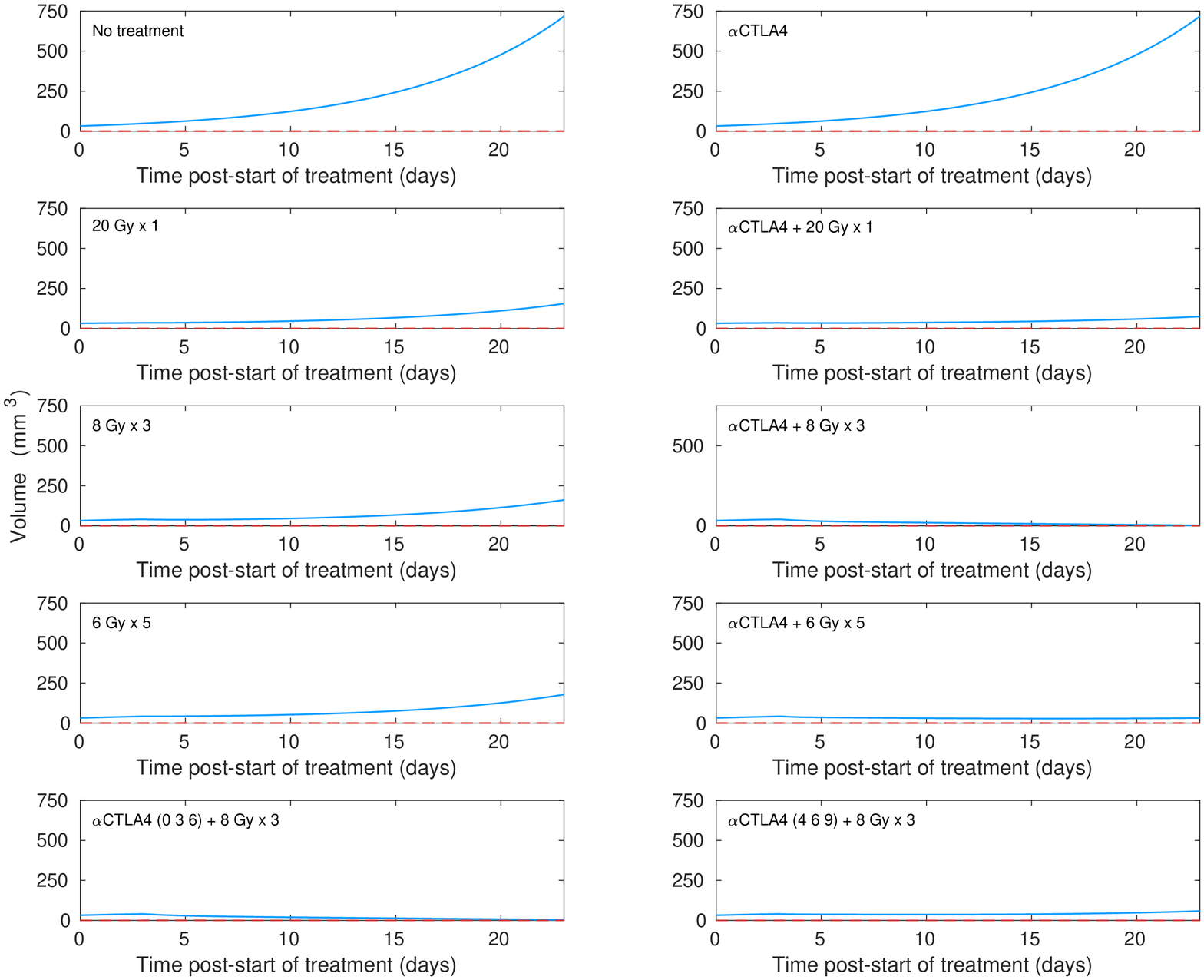}
	\caption{Contribution of tumor cells (blue solid lines) and T-cells (red dashed lines) to tumor volumes in fits of our biomathematical model to experimental data reported by Dewan et al. \cite{dewan2009}. These curves are obtained from fits reported in Fig.~\ref{fig_DewanFitting2} in the main article, where we used the classical LQ-model to account for direct cell death and the effect of vascular damage was taken into account. Tumor volumes are dominated by tumor cells.}
	\label{}
\end{figure*}

\newpage
\begin{figure*}[h!]
	\renewcommand{\arraystretch}{1.3}
	\centering
	\includegraphics[width=6in]{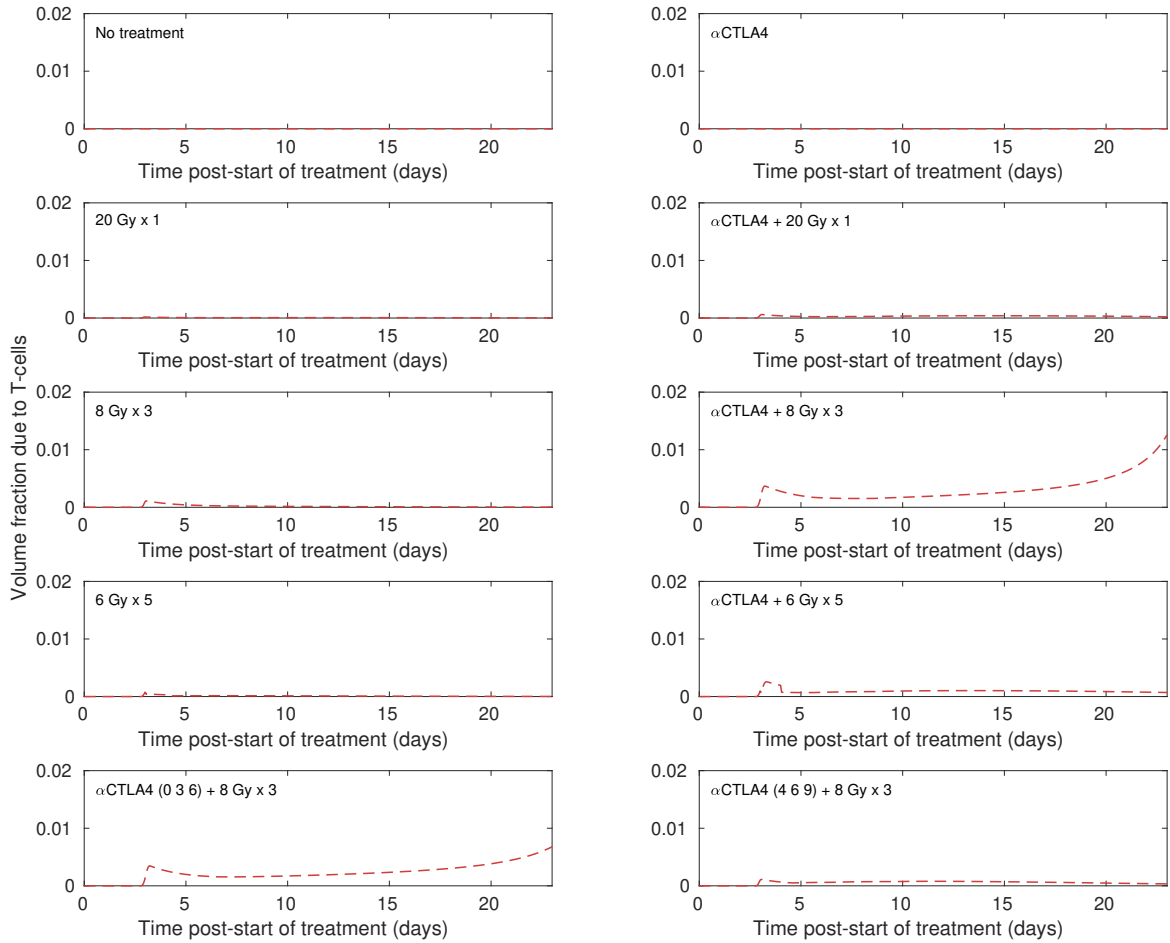}
	\caption{Relative contribution of T-cells (red dashed lines) to tumor volumes in fits of our biomathematical model to experimental data reported by Dewan et al. \cite{dewan2009}. These curves are obtained from fits reported in Fig.~\ref{fig_DewanFitting2} in the main article, where we used the classical LQ-model to account for direct cell death and the effect of vascular damage was taken into account. Tumor volumes are dominated by tumor cells, and only when the tumor is close to remission the fraction of T-cells becomes large.}
	\label{}
\end{figure*}
\newpage

\section{Supplementary Tables}
\begin{table*}[!h]
	\renewcommand{\arraystretch}{1.3}
	\caption{Best-fitting parameters of our model to experimental data reported in \cite{dewan2009} and \cite{deng2014}. A modified LQ-model was used to account for tumor cell radiosensitivity (Eq. (1) and (2)). The symbol~* indicates that the parameter value was fixed and not included in the  optimization, while~** indicates that the parameter was optimized, but the same value was used for $\alpha$CTLA4 and $\alpha$PDL1 fitting. Dewan and Deng refer to best-fitting parameters to data reported in \cite{dewan2009} and \cite{deng2014}, respectively. Initial numbers/concentrations of variables not indicated in the table ($C_\mathrm{d}$, $\hat{A}$, $T_\mathrm{a}$, $\hat{T}_\mathrm{a}$, $\hat{T}_\mathrm{b}$, $c_4$, $p_1$) are set to zero.}
	\label{tab_param}
	\centering
	\resizebox{15cm}{!} {
		\begin{tabular}{|c|c|c|}
			\hline
			\multicolumn{3}{|c|}{\bfseries Parameters/Initial conditions} \\
			\hline
			\bfseries Parameter & \bfseries Description [units] & \bfseries Value\\
			\hline\hline
			$C (t=0)$ & *Initial number of tumor cells & 100\\
			\hline
			$\hat{T} (t=0)$ & **Initial pool of \textit{blank} T-cells in activation site & $1.3188\times10^6$\\
			\hline
			$\lambda_1$ & Proliferation rate of tumor cells [day$^{-1}$] & $0.1367$ (Dewan); $0.1217$ (Deng)\\
			\hline
			$\lambda_2$ & Saturation of proliferation & $0$ (Dewan); $4.5320\times10^{-10}$ (Deng)\\
			\hline
			$\alpha_\mathrm{C}$ & Linear parameter of LQ model for tumor cells [Gy$^{-1}$] & $0.0200$ (Dewan); $0.0299$ (Deng)\\
			\hline
			$\beta_\mathrm{C}$ & Quadratic parameter of LQ model for tumor cells [Gy$^{-2}$] & $0.0068$ (Dewan); $0.00299$ (Deng*)\\
			\hline
			$c$ & Modulation factor of the $\beta_\mathrm{C}$ parameter [Gy$^{-1/2}$] & $-0.2001$ (Dewan); $0$ (Deng*)\\
			\hline
			$\phi$ & **Rate of elimination of doomed tumor cells [day$^{-1}$] & $0.0484$\\
			\hline
			${\tau_\mathrm{d}}_1$ & *Mitotic delay (start) [day] & $1$\\
			\hline
			${\tau_\mathrm{d}}_2$ & *Mitotic delay (end) [day] & $1.5000$\\
			\hline
			$p$ & **Immune-mediated tumor cell death rate [day$^{-1}$] & $24.4339$\\
			\hline
			$q$ & **Immune-mediated tumor cell death parameter & $0.6925$\\
			\hline
			$s$ & **Slope of the immune-mediated tumor death curve & $7.5013$\\
			\hline
			$i_{\mathrm{p}_1}$ & Dose of $\alpha$PDL1 per fraction & $19.0387$ (if $\alpha$PDL1) $0$ (otherwise)\\
			\hline
			$\mu$ & *Rate of elimination of $\alpha$PDL1 [day$^{-1}$] & $0.5000$\\
			\hline
			$\rho$ & **Rate of natural production of antigens [day$^{-1}$] & $1.0222\times10^{-19}$\\
			\hline
			$\psi$ & **Rate of release of antigens [day$^{-1}$] & $267.2069$\\
			\hline
			$\tau_1$ & **Delay between antigen liberation and T-cell activation [day] & $0.9894$\\
			\hline
			$\sigma$ & **Rate of natural elimination of antigens [day$^{-1}$] & $0.0300$\\
			\hline
			$a$ & **Rate of activation of T-cells [day$^{-1}$] & $8.8379\times10^{-8}$\\
			\hline
			$r$ & *Ratio of activation/inactivation rates & $5$\\
			\hline
			$i_{\mathrm{c}_4}$ & Dose of $\alpha$CTLA4 per fraction & $5.0564$ (if $\alpha$CTLA4) 0 (otherwise)\\
			\hline
			$\nu$ & *Rate of elimination of $\alpha$CTLA4 [day$^{-1}$] & $0.1000$\\
			\hline
			$\tau_2$ & **Time taken by activated T-cell to reach the tumor [day] & $0.4561$\\
			\hline
			$\alpha_\mathrm{T}$ & *Linear parameter of LQ model for T-cells [Gy$^{-1}$] & $0.1000$\\
			\hline
			$\beta_\mathrm{T}$ & *Quadratic parameter of LQ model for T-cells [Gy$^{-2}$] & $0.0100$\\
			\hline
			$\iota$ & **T-cells death rate due to interaction with tumor cells [day$^{-1}$] & $1.9847\times10^{-8}$\\
			\hline
			$\eta$ & **Rate of natural elimination of T-cells [day$^{-1}$] & $0.0328$\\
			\hline
			$h$ & **Rate of production of \textit{blank} T-cells [day$^{-1}$] & $1.7391\times10^{5}$\\
			\hline
			$V_\mathrm{C}$ & *Volume of tumor cells [mm$^3$] & $10^{-6}$\\
			\hline
			$V_\mathrm{T}$ & *Volume of T-cells [mm$^3$] & $2\times10^{-7}$\\
			\hline
		\end{tabular}
	}
\end{table*}

\newpage

\begin{table*}[!h]
	\renewcommand{\arraystretch}{1.3}
	\caption{List of best-fitting parameters of our model to experimental data reported in reference \cite{dewan2009}. The LQL model was used to account for tumor cell radiosensitivity (Eq. (3)). The symbol~* indicates that the parameter value was fixed and not included in the optimization. Initial numbers/concentrations of variables not indicated in the table ($\mathit{C_\mathrm{d}}$, $\hat{A}$, $T_\mathrm{a}$, $\mathit{\hat{T}_\mathrm{a}}$, $\mathit{\hat{T}_\mathrm{b}}$, $c_4$, $p_1$) are set to zero.}
	\label{tab_param_vd}
	\centering
	\resizebox{15cm}{!} {
	\begin{tabular}{|c|c|c|}
		\hline
		\multicolumn{3}{|c|}{\bfseries Parameters/Initial conditions} \\
		\hline
		\bfseries Parameter & \bfseries Description [units] & \bfseries Value\\
		\hline\hline
		$C (t=0)$ & *Initial number of tumor cells & 100\\
		\hline
		$\hat{T} (t=0)$ & Initial pool of \textit{blank} T-cells in activation site & $1.8784\times10^6$\\
		\hline
		$\lambda_1$ & Proliferation rate of tumor cells [day$^{-1}$] & $0.1356$\\
		\hline
		$\lambda_2$ & Saturation of proliferation & $9.1172\times10^{-118}$\\
		\hline
		$\alpha_\mathrm{C}$ & Linear parameter of LQ model for tumor cells [Gy$^{-1}$] & $0.0442$\\
		\hline
		$\beta_\mathrm{C}$ & Quadratic parameter of LQ model for tumor cells [Gy$^{-2}$] & $0.0167$\\
		\hline
		$x$ & Modulation factor of the $\beta_\mathrm{C}$ parameter [Gy$^{-1}$] & $8.4298$\\
		\hline
		$\phi$ & Rate of elimination of doomed tumor cells [day$^{-1}$] & $0.0312$\\
		\hline
		${\tau_\mathrm{d}}_1$ & *Mitotic delay (start) [day] & $1$\\
		\hline
		${\tau_\mathrm{d}}_2$ & *Mitotic delay (end) [day] & $1.5000$\\
		\hline
		$p$ & Immune-mediated tumor cell death rate [day$^{-1}$] & $23.3045$\\
		\hline
		$q$ & Immune-mediated tumor cell death parameter & $0.7212$\\
		\hline
		$s$ & Slope of the immune-mediated tumor death curve & $8.8282$\\
		\hline
		$\rho$ & Rate of natural production of antigens [day$^{-1}$] & $1.6009\times10^{-31}$\\
		\hline
		$\psi$ & Rate of release of antigens [day$^{-1}$] & $9.7669\times10^{3}$\\
		\hline
		$\tau_1$ & Delay between antigen liberation and T-cell activation [day] & $2.5494$\\
		\hline
		$\sigma$ & Rate of natural elimination of antigens [day$^{-1}$] & $0.2905$\\
		\hline
		$a$ & Rate of activation of T-cells [day$^{-1}$] & $2.8156\times10^{-7}$\\
		\hline
		$r$ & *Ratio of activation/inactivation rates & $5$\\
		\hline
		$i_{\mathrm{c}_4}$ & Dose of $\alpha$CTLA4 per fraction & $1.0865\times10^{3}$\\
		\hline
		$\nu$ & *Rate of elimination of $\alpha$CTLA4 [day$^{-1}$] & $0.1000$\\
		\hline
		$\tau_2$ & Time taken by activated T-cell to reach the tumor [day] & $0.2257$\\
		\hline
		$\alpha_\mathrm{T}$ & *Linear parameter of LQ model for T-cells [Gy$^{-1}$] & $0.1000$\\
		\hline
		$\beta_\mathrm{T}$ & *Quadratic parameter of LQ model for T-cells [Gy$^{-2}$] & $0.0100$\\
		\hline
		$\iota$ & Rate of death of T-cells due to interaction with tumor cells [day$^{-1}$] & $2.2922\times10^{-8}$\\
		\hline
		$\eta$ & Rate of natural elimination of T-cells [day$^{-1}$] & $0.2191$\\
		\hline
		$h$ & Rate of production of \textit{blank} T-cells [day$^{-1}$] & $2.9738\times10^{5}$\\
		\hline
		$V_\mathrm{C}$ & *Volume of tumor cells [mm$^3$] & $10^{-6}$\\
		\hline
		$V_\mathrm{T}$ & *Volume of T-cells [mm$^3$] & $2\times10^{-7}$\\
		\hline
	\end{tabular}}
\end{table*}

\newpage

\begin{table*}[!h]
	\renewcommand{\arraystretch}{1.3}
	\caption{List of best-fitting parameters of our model to experimental data reported in reference \cite{dewan2009}. The LQ model was used to account for tumor cell radiosensitivity (Eq. (1)), and limited T-cell infiltration was assumed for 20~Gy irradiation due to vascular damage (Eq. (15)). The symbol~* indicates that the parameter value was fixed and not included in the optimization. Initial numbers/concentrations of variables not indicated in the table ($C_\mathrm{d}$, $\hat{A}$, $T_\mathrm{a}$, $\hat{T}_\mathrm{a}$, $\hat{T}_\mathrm{b}$, $c_4$, $p_1$) are set to zero.}
	\label{tab_param_LQL}
	\centering
	\resizebox{15cm}{!} {
	\begin{tabular}{|c|c|c|}
		\hline
		\multicolumn{3}{|c|}{\bfseries Parameters/Initial conditions} \\
		\hline
		\bfseries Parameter & \bfseries Description [units] & \bfseries Value\\
		\hline\hline
		$C (t=0)$ & *Initial number of tumor cells & 100\\
		\hline
		$\hat{T} (t=0)$ & Initial pool of \textit{blank} T-cells in activation site & $1.1679\times10^6$\\
		\hline
		$\lambda_1$ & Proliferation rate of tumor cells [day$^{-1}$] & $0.1357$\\
		\hline
		$\lambda_2$ & Saturation of proliferation & $3.3764\times10^{-109}$\\
		\hline
		$\alpha_\mathrm{C}$ & Linear parameter of LQ model for tumor cells [Gy$^{-1}$] & $0.0200$\\
		\hline
		$\beta_\mathrm{C}$ & Quadratic parameter of LQ model for tumor cells [Gy$^{-2}$] & $0.0022$\\
		\hline
		$\phi$ & Rate of elimination of doomed tumor cells [day$^{-1}$] & $0.0300$\\
		\hline
		${\tau_\mathrm{d}}_1$ & *Mitotic delay (start) [day] & $1$\\
		\hline
		${\tau_\mathrm{d}}_2$ & *Mitotic delay (end) [day] & $1.5000$\\
		\hline
		$p$ & Immune-mediated tumor cell death rate [day$^{-1}$] & $24.9382$\\
		\hline
		$q$ & Immune-mediated tumor cell death parameter & $0.6886$\\
		\hline
		$s$ & Slope of the immune-mediated tumor death curve & $5.7249$\\
		\hline
		$\rho$ & Rate of natural production of antigens [day$^{-1}$] & $1.3746\times10^{-27}$\\
		\hline
		$\psi$ & Rate of release of antigens [day$^{-1}$] & $3.4596\times10^7$\\
		\hline
		$\tau_1$ & Delay between antigen liberation and T-cell activation [day] & $1.0806$\\
		\hline
		$\sigma$ & Rate of natural elimination of antigens [day$^{-1}$] & $0.1356$\\
		\hline
		$a$ & Rate of activation of T-cells [day$^{-1}$] & $6.8010\times10^{-12}$\\
		\hline
		$r$ & *Ratio of activation/inactivation rates & $5$\\
		\hline
		$i_{\mathrm{c}_4}$ & Dose of $\alpha$CTLA4 per fraction & $7.4987$\\
		\hline
		$\nu$ & *Rate of elimination of $\alpha$CTLA4 [day$^{-1}$] & $0.1000$\\
		\hline
		$\tau_2$ & Time taken by activated T-cell to reach the tumor [day] & $0.6077$\\
		\hline
		$\alpha_\mathrm{T}$ & *Linear parameter of LQ model for T-cells [Gy$^{-1}$] & $0.100$\\
		\hline
		$\beta_\mathrm{T}$ & *Quadratic parameter of LQ model for T-cells [Gy$^{-2}$] & $0.0100$\\
		\hline
		$\iota$ & Rate of death of T-cells due to interaction with tumor cells [day$^{-1}$] & $6.2472\times10^{-9}$\\
		\hline
		$\eta$ & Rate of natural elimination of T-cells [day$^{-1}$] & $0.5031$\\
		\hline
		$h$ & Rate of production of \textit{blank} T-cells [day$^{-1}$] & $1.3358\times10^{5}$\\
		\hline
		$V_\mathrm{C}$ & *Volume of tumor cells [mm$^3$] & $10^{-6}$\\
		\hline
		$V_\mathrm{T}$ & *Volume of T-cells [mm$^3$] & $2\times10^{-7}$\\
		\hline
	\end{tabular}}
\end{table*}
\newpage

\begin{table*}[!h]
	\renewcommand{\arraystretch}{1.3}
	\caption{Parameter restrictions used for model fitting. Parameters not included in the table have a positivity constraint.}
	\label{tab_param_const}
	\centering
	\resizebox{15cm}{!} {
		\begin{tabular}{|c|c|c|}
			\hline
			\multicolumn{3}{|c|}{\bfseries Parameters/Initial conditions constraints} \\
			\hline
			\bfseries Parameter & \bfseries Description [units] & \bfseries Bound\\
			\hline\hline
			$\alpha_\mathrm{C}$ & Linear parameter of LQ model for tumor cells [Gy$^{-1}$] & $[0.02, 0.35]$\\
			\hline
			$\beta_\mathrm{C}$ & Quadratic parameter of LQ model for tumor cells [Gy$^{-2}$] & $[\alpha_\mathrm{C}/20, \alpha_\mathrm{C}/2]$\\
			\hline
			$c$ & Modulation factor of $\beta_\mathrm{C}$ in the modified LQ-model [Gy$^{-1/2}$] & $[-0.2236, \infty)$\\
			\hline
			$\phi$ & Rate of elimination of doomed tumor cells [day$^{-1}$] & $[0.03, 0.7]$\\
			\hline
			$q$ & Immune-mediated tumor cell death parameter & $[0, 1]$\\
			\hline
			$\tau_1$ & Delay between antigen liberation and T-cell activation [day] & $[0, 5]$\\
			\hline
			$\sigma$ & Rate of natural elimination of antigens [day$^{-1}$] & $[0.03, 0.7]$\\
			\hline
			$a$ & Rate of activation of T-cells [day$^{-1}$] & $[0, 3\times10^{-7}]$\\
			\hline
			$\tau_2$ & Time taken by activated T-cell to reach the tumor [day] & $[0, 5]$\\
			\hline
			$\iota$ & T-cells death rate due to interaction with tumor cells [day$^{-1}$] & $[0, 10^{-7}]$\\
			\hline
			$\eta$ & Rate of natural elimination of T-cells [day$^{-1}$] & $[0.03, 0.7]$\\
			\hline
			$h$ & Rate of production of \textit{blank} T-cells [day$^{-1}$] & $[0.05 \hat{T}_0, \hat{T}_0]$\\
			\hline
		\end{tabular}
	}
\end{table*}

\newpage

\section{Supplementary Information}
\subsection{TCP calculation}
\subsubsection{Markov Stochastic Model}

We have employed the \textit{clonogenic cell hypothesis} \cite{nahum1993} to obtain tumor control probablities (TCP) from our model, which states that in order to control the tumor, all cells with proliferative capacity (which we identify with the compartment ${{C}}$) need to be eliminated. In order to perform the calculation of TCPs, we implemented a Markov birth/death stochastic process \cite{hanin2001} by interpreting terms in the differential equations as birth/death probabilities. In order to limit computation times, the stochastic Markov only comes into operation when the number of undamaged tumor cells falls below a certain threshold, in this case 1000 cells.

For each cell, a probabilistic experiment is performed to decide whether the cell multiplies, dies, or none, resulting in 2, 0 or 1 cells, as sketched in Supplementary Fig.~\ref{fig_markov}. The probability of proliferation, $P_\mathrm{p}$, and death, $P_\mathrm{d}$, are given by the corresponding terms in the model:
\begin{eqnarray}
	\displaystyle
	P_\mathrm{p} &=& \lambda_1 \left(1-\lambda_2{{C_\mathrm{tot}}}(t)\right) \Delta t \nonumber\\
	\displaystyle
	P_\mathrm{d} &=&  p (1+p_1(t))\frac{({{T_\mathrm{a}}}(t)/{{C_\mathrm{tot}}}(t))^q}{s+({{T_\mathrm{a}}}(t)/{{C_\mathrm{tot}}}(t))^q} \Delta t \nonumber
\end{eqnarray}
where $\Delta t$ is the discretization time step.

The respective code of the Markov model is available on the Dataverse repository~\cite{glezCrespo2019} and can be seen within the function called \texttt{markov\_TCP\_analysis} (\texttt{radioimmuno\_response\_model} function file). It is summarized in that for each cell the resulting number of cells is:
\texttt{randsrc(1,1,[2,0,1; Pp,Pd,1-Pp-Pd])}.
\vspace{0.3cm}
\begin{figure}[h]
	\renewcommand{\arraystretch}{1.3}
	\centering
	\includegraphics[width=1in]{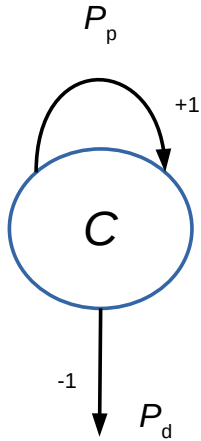}
	\caption{Sketch of the process described by the Markov model.}
	\label{fig_markov}
\end{figure}

\subsubsection{Parameters Perturbations}

In addition to the stochasticity provided by the Markov model, variations of the best-fitting parameters are introduced to simulate the response of a population different subjects to the treatment. To obtain our results, we performed 200 simulations with random normal perturbations with a 5\% standard deviation with around best-fitting parameter values (only those that have not been fixed). This is coded in the function called \texttt{calculate\_TCP}.

The TCP of the population is then calculated as the ratio of successful cases (where success means that no undamaged tumour cells are left alive, i.e. $C=0$) to the total number of experiments (200).

\newpage
\subsection{Explicit Euler Method}
Given the demands in terms of execution time required by the optimizations performed in this work, an explicit Euler method has been implemented. The Euler method may present stability problems in \textit{stiff} differential equations, but in general it is not possible to know \textit{a priori} whether a given equation will present \textit{stiffness}. To ensure the stability and convergence of the method we have compared the solutions obtained for different time steps. Supplementary Fig.~\ref{fig_dt} shows the evolution of tumor volumes (obtained with best-fitting parameters reported in Supplementary Table \ref{tab_param_const}) with no treatment (left) and with $8~\mathrm{Gy}\times3$ and $\alpha$CTLA4 at days 2, 5 and 8 (right) for different discretizations, ranging from 0.5 days to 0.001 days). Supplementary Fig.~\ref{fig_error} presents the absolute error made when calculating the volume with each of the time steps in relation to the solution obtained for the smallest of them, taking day 15 as a reference. The error has a linear dependence on the discretization step, as expected for the Euler method in well-conditioned problems. Using $\Delta t=0.05~\mathrm{day}$ provides a good tradeoff between fast computation and qualitatively accurate solutions (especially when considering the uncertainties of the experimental data).
\begin{figure}[h]
	\renewcommand{\arraystretch}{1.3}
	\centering
	\includegraphics[width=4.5in]{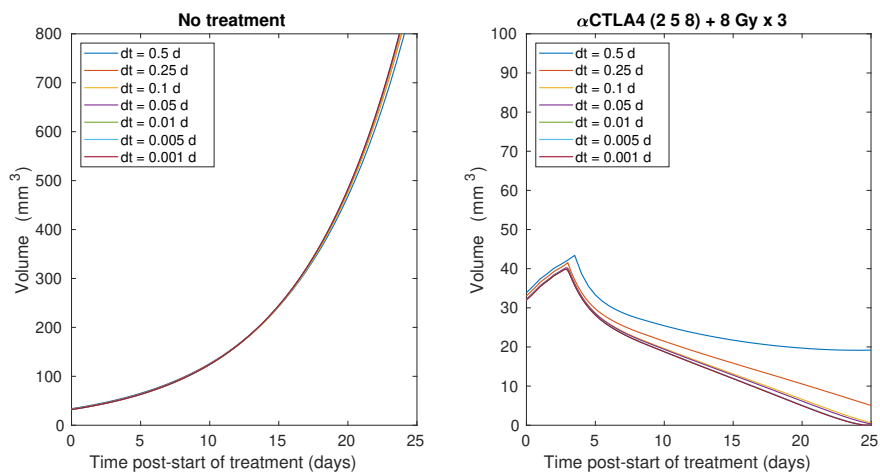}
	\caption{Tumor volume evolution with different time steps.}
	\label{fig_dt}
\end{figure}
\begin{figure}[h]
	\renewcommand{\arraystretch}{1.3}
	\centering
	\includegraphics[width=4.5in]{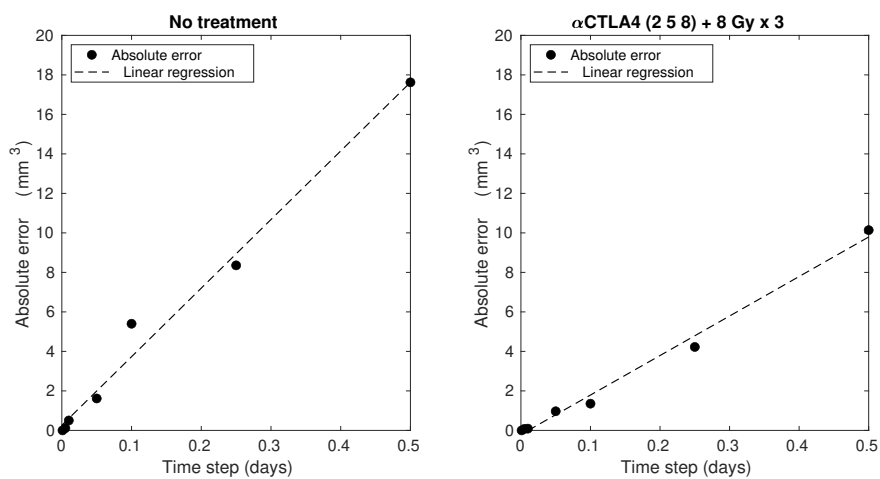}
	\caption{Absolute error when calculating the volume (at day 15 post-start of treatment) with different time steps (black circles), and linear fits of the values obtained (dashed line). The solution obtained with the smallest time step (0.001 days) serves as reference for calculating the errors.}
	\label{fig_error}
\end{figure}

\end{document}